\documentclass[prl,twocolumn,floatfix,superscriptaddress]{revtex4-2}

\usepackage{amsmath, amssymb}
\usepackage{mathtools}
\usepackage{times}
\usepackage{graphicx}
\usepackage{bm}
\usepackage{xspace}
\usepackage{siunitx}
\usepackage{xcolor}
\usepackage{braket}
\usepackage{comment}
\usepackage{hyperref}

\definecolor{myColor}{rgb}{0.02,0.12,0.8}
\definecolor{myciteColor}{rgb}{0.39,1.0,0.89}
\DeclareSymbolFont{extraup}{U}{zavm}{m}{n}
\DeclareMathSymbol{\varheart}{\mathalpha}{extraup}{86}
\DeclareMathSymbol{\vardiamond}{\mathalpha}{extraup}{87}

\DeclareSIUnit{\G}{G}
\renewcommand{\figurename}[1]{Fig.~}

\newcommand\identity{1\kern-0.25em\text{l}}

\begin{document}

\title{Polarization- and wave-vector selective optical metasurface with near-field coupling}
\author{Helene Wetter}
\email{helene.wetter@uni-paderborn.de}
\author{Jan Wingenbach}
\author{Falk Rehberg}
\affiliation{Department of Physics, Paderborn University, Warburger Str. 100, 33098 Paderborn, Germany}
\author{Wenlong Gao}
\affiliation{Eastern Institute of Technology, Eastern Institute for Advanced Study, No. 2911 Haijiang Avenue, Ningbo, 315200, China}
\author{Stefan Schumacher}
\author{Thomas Zentgraf}
\affiliation{Department of Physics, Paderborn University, Warburger Str. 100, 33098 Paderborn, Germany}
\affiliation{Institute for Photonic Quantum Systems (PhoQS), Paderborn University, Warburger Str. 100, 33098 Paderborn, Germany}

\date{\today}

\begin{abstract}
Metasurfaces are a powerful tool for manipulating light using small structures on the nanoscale. In most metasurfaces, near-field couplings are treated as unfavorable perturbations. Here, we experimentally investigate a structure consisting of sinusoidally modulated silicon waveguides where near-field coupling of local resonances leads to negative coupling, i.e.~a negative coupling constant.
This gives rise to wave-vector dependent eigenstates of elliptical, linear and circular polarizations. In particular, fully circular polarization states are not only present at a single point in momentum-space ($k$-space), but along a line. 
This circular polarization line, as well as a linear polarization line, emanates from a polarization degeneracy at the Dirac point. We experimentally validate the existence of these eigenstates and demonstrate the energy-, polarization- and wave-vector-dependence of this metasurface.
By tuning the incident $k$-vector, certain polarization-energy eigenstates are strongly reflected allowing for uses in angle-tunable polarization filters and light sources. 
\end{abstract}
\maketitle
\section{Introduction}   
In the last decades, a wide variety of metasurfaces have been designed and analyzed, enabling precise control of light with sub-micrometer-thin planar structures. These engineered platforms provide a broad range of applications at optical frequencies ranging from beam steering \cite{yu2011light, spinelli2012broadband, geromel2023geometric} and holography \cite{zheng2015metasurface, huang2013three} to nonlinear optics \cite{klein2006second,zhang2013coherent,li2017nonlinear}. The compact and flexible design provides advantages in integrated photonics \cite{meng2021optical}, imaging \cite{ou2023advances} and sensing \cite{georgi2019metasurface, kim_chiroptical_2021}. 
Commonly, metasurfaces are composed of
meta-atoms that locally modify the amplitude and phase of the incident electric field according to their shape, size and rotation angle. Multiple meta-atoms are then arranged within an array such that the local interactions add up to the intended global behavior, for example refraction \cite{yu2011light}.
Using chiral meta-atoms i.e.~geometries that cannot be replicated by its mirrored image, metasurfaces can provide chiroptical effects such as circular dichroism, where the absorption coefficient for right- and left circular polarization differs \cite{schaferling2012tailoring, mandal2018large, zhu2018giant, hu2017all, khaliq_recent_2023}. 
For such metasurfaces, the near-field coupling of the meta-atoms is mostly neglected or treated as an unfavorable effect if a highly local response is desired. 

In contrast, metasurfaces that exploit coupling between meta-atoms, such as those realizing (quasi-) bound states in the continuum (BICs), use collective modes to achieve high-Q resonances and enhance the performance of metasurfaces beyond traditional designs \cite{shi_planar_2022, zentgraf2004tailoring,kodigala2017lasing,koshelev2020subwavelength,christ2006controlling}.
Chiral metasurfaces supporting \mbox{(quasi-)}BICs can exhibit vortex polarization singularities that can be decomposed into circularly polarized states (C points) and lines of linearly polarized states (L line) in momentum-space \cite{shi_planar_2022,liu_circularly_2019}. 
Gao et al.~proposed a metasurface design that obeys unique wave-vector dependent circularly polarized states that are not only present at one point in the momentum-space (C point) but along a line (C line) \cite{gao2022spin}. Such design can expand the capabilities of chiral metasurfaces 
and opens up avenues for applications in spin-selective sensing. The key concept is based on near-field coupled local guided-mode resonances that lead to negative coupling. 
Recently, similar guided-mode resonances, that weakly couple to incident light, were used for electro-optical wavefront shaping \cite{barton2021wavefront, lawrence2020high}.

Here, we experimentally demonstrate and verify the proposed theoretical concept of wave-vector dependent polarization eigenstates by coupled local guided-mode resonances. 
The metasurface is expected to exhibit polarization-selective and wave-vector dependent radiative eigenstates, including both linear and circular polarizations along a line in momentum-space ($k$-space).
When illuminating the metasurface, the light matching the polarization and frequency of the wave-vector dependent eigenstate gets reflected. This allows the metasurface to generate light with a desired polarization and frequency by adressing the appropriate angle in momentum-space. 

\section{Basic principle}  
\begin{figure*}[t!]
    \centering
    \includegraphics[width=\textwidth]{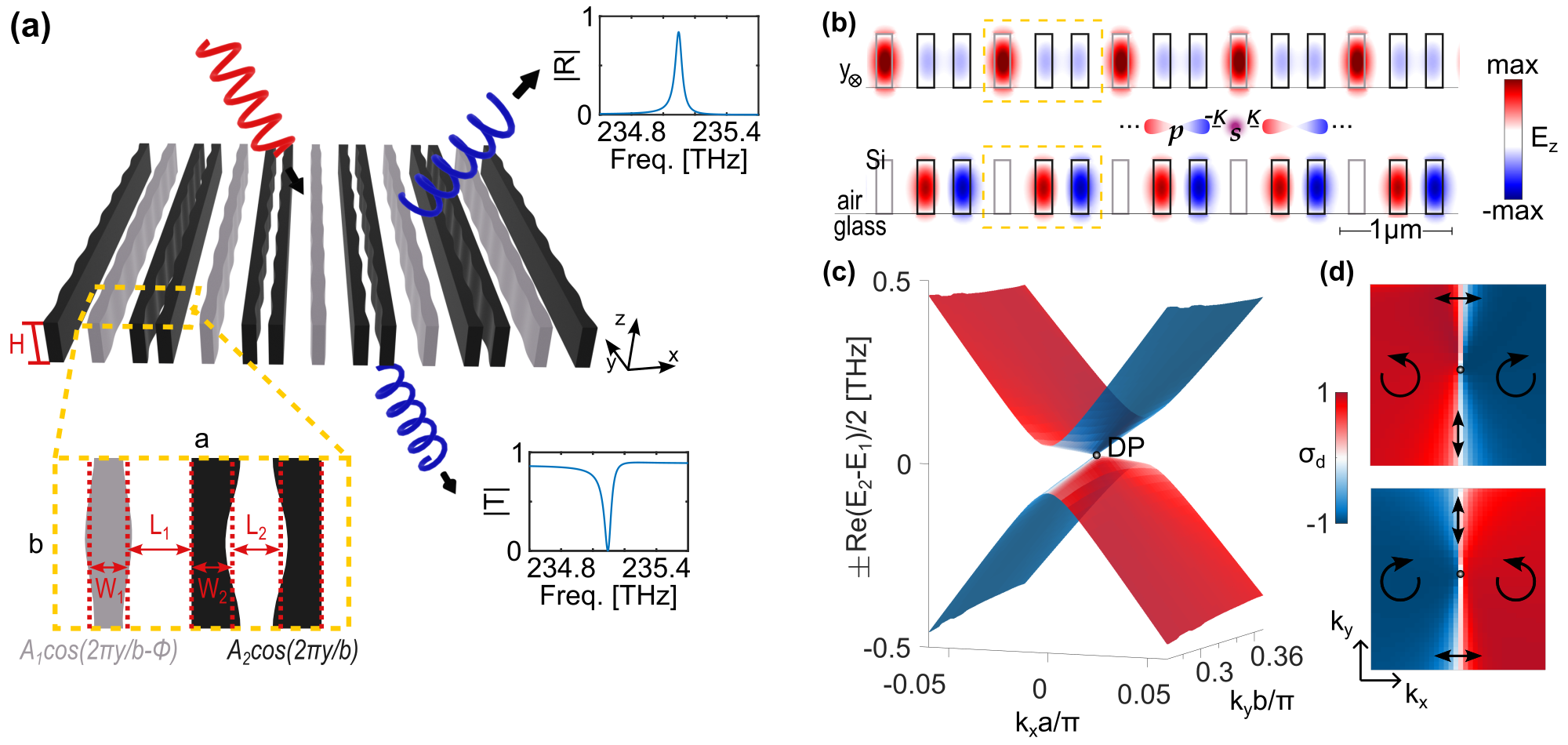}
\caption{(a)~Sketch of the discussed metasurface composed of single waveguide (SW - gray) and double waveguide (DW - black). A top view of the unit cell with all its geometrical parameters is shown in the inset/ yellow box. At resonance frequency, the incident light is reflected (see reflection $|R|$) leading to a dip in the transmission spectrum ($|T|$). Because of an additional phase shift, 
  the reflected and transmitted light is cross-circularly polarized. (b)~Side view of the unmodulated silicon waveguides on glass and mode profile ($E_z$ component) of the fundamental modes. Top: TE-like mode in the SW, Bottom: odd mode in DW. (c)~Real difference of eigenenergies $E_1$ and $E_2$ in the momentum-space ($k$-space) around the Dirac point (DP). The color indicates the eigenpolarization $\sigma_d$ which is shown in (d) for the upper and lower surface separately. The black arrows illustrate the polarization state.}
    \label{fig:1}
\end{figure*} 
%
The aforementioned metasurface structure consists of sinusoidally modulated single and double silicon waveguides and is sketched in Figure \ref{fig:1}(a). The unit cell (UC) can be described by a single waveguide (SW - gray) of medium width $W_1$ that is modulated sinusoidally on both sides with the amplitude $A_1$ over the period of $b$ with additional phase shift $\Phi$. The double waveguide (DW - black) consists of two waveguides at distance $L_2$ where only the edges facing each other are sinusoidally modulated with amplitude $A_2$. The medium width of one DW is $W_2$ and it's distance to the SW is given by $L_1$. All waveguides are equal in height $H$ and located on top of a glass substrate. 

To understand the behavior of this metasurface, we focus first on the unmodulated waveguides ($A_1=A_2=0$).  %
This waveguide structure provides three fundamental modes: A TE-like mode in the SW and an odd and even mode in the DW~\cite{gao2022spin}. The mode profile of the first two is shown in Figure \ref{fig:1}(b). 
The waveguide parameters are set to 
$W_1=\SI{141}{\nano\meter}$, $W_2=\SI{150}{\nano\meter}$, $L_1=\SI{220}{\nano\meter}$, $L_2=\SI{171.5}{\nano\meter}$ and $H=\SI{478}{\nano\meter}$
such that the dispersion of the two individual waveguide modes intersects at the desired operating frequency (\SI{240}{\tera\hertz}) of the metasurface and coupling of the two modes occurs. 

Considering the field distribution of the modes depicted in Fig.~\ref{fig:1}(b), the SW mode (top) resembles an s-orbit, while the odd DW mode (bottom) resembles a p-orbit (refer to illustration of orbits in the center of Fig.~\ref{fig:1}(b)).
Thus, the mode coupling can be understood as an inter-orbit coupling, which can lead to negative coupling. Here, the coupling constant $\kappa$ between the s-orbit and its neighboring p-orbits is of opposite sign for the left and right neighbor. 
This alternating positive and negative coupling can be explained by the field distribution within the waveguides (see Fig.~\ref{fig:1}(b)): On the left side of the DW, the field of the DW mode is in phase with the SW mode; on the right side they are phase-inverted which leads to a coupling factor $\kappa$ of opposite sign. 

Applying the alternating sign of the coupling constant into the tight-binding model leads to the eigenstates $\bra{+}=[1;i]^T$ and $\bra{-}=[1;-i]^T$ (more details are given in \cite{gao2022spin}), where the first (second) component of the vector describes the photon state in the SW (DW). This means, that for the coupled modes, the phase difference between the SW and DW is $\pm\frac{\pi}{2}$, which can construct circular polarization states in the far field.
It is worth noting, that the unmodulated waveguides add a phase to the transmitted and reflected light because the effective reflective index is different in $x$ and $y$.
The geometry of the discussed metasurface is designed to add a phase shift of $\pi$ to the $y$-component for normal incidence such that the helicity of the circular polarization is changed. 

The sinusoidal modulation is added to the structure to allow the eigenstates to couple to the ambient environment i.e.~allow the metasurface to interact with incident light. A top view of the resulting unit cell is depicted in Figure~\ref{fig:1}(a)(yellow inset).
For tailored amplitudes $A_1=\SI{10}{\nano\meter}$, $A_2=\SI{18.2}{\nano\meter}$, period $b=\SI{660}{\nano\meter}$ and the phase $\Phi=\SI{0.25}{}$ of the sinusoidal modulation, the structure features a Dirac point (DP) in momentum-space (Fig.~\ref{fig:1}(c)). Figure \ref{fig:1}(c) depicts the
real difference of eigenenergies $E_1$ and $E_2$ centered around \SI{0}{\tera\hertz} for a better visualization of the Dirac cone. 
The Dirac point coordinate will be denoted as ($K_x, K_y$). 
Besides the eigenenergies, the eigenpolarizations are color coded in Figure \ref{fig:1}(c) and (d) by the polarization parameter $\sigma_d$ given by
\begin{equation}
    \sigma_d=\mathrm{Im}\Big(\frac{\vec{E^*} \times \vec{E}}{|\vec{E}|^2}\Big)\cdot\frac{\vec{k}}{|\vec{k}|}
\end{equation}
with $\vec{E}$ being the electric field and $\vec{k}$ the wavevector. Thus, $\sigma_d=1$ ($-1$) means left (right) circular polarization and $\sigma_d=0$ is associated with linear polarization of unknown polarization axis. Everything in between corresponds to elliptical polarization. 
Figure \ref{fig:1}(d) images the eigenpolarization $\sigma_d$ for the upper and lower energy level. It shows, that circling around the Dirac point (DP) covers both linear and circular polarization states, and that the eigenpolarizations of the upper and lower energy levels are mirrored along $k_x=0$. 
The linear polarization states are aligned along the $k_y$-axis, while the circular polarization states extend along a line parallel to the $k_x$-axis through the Dirac point.
The Dirac point itself features a polarization degeneracy.

\section{Theoretical expectations}  
\begin{figure}[]
    \centering
    \includegraphics[width=\columnwidth]{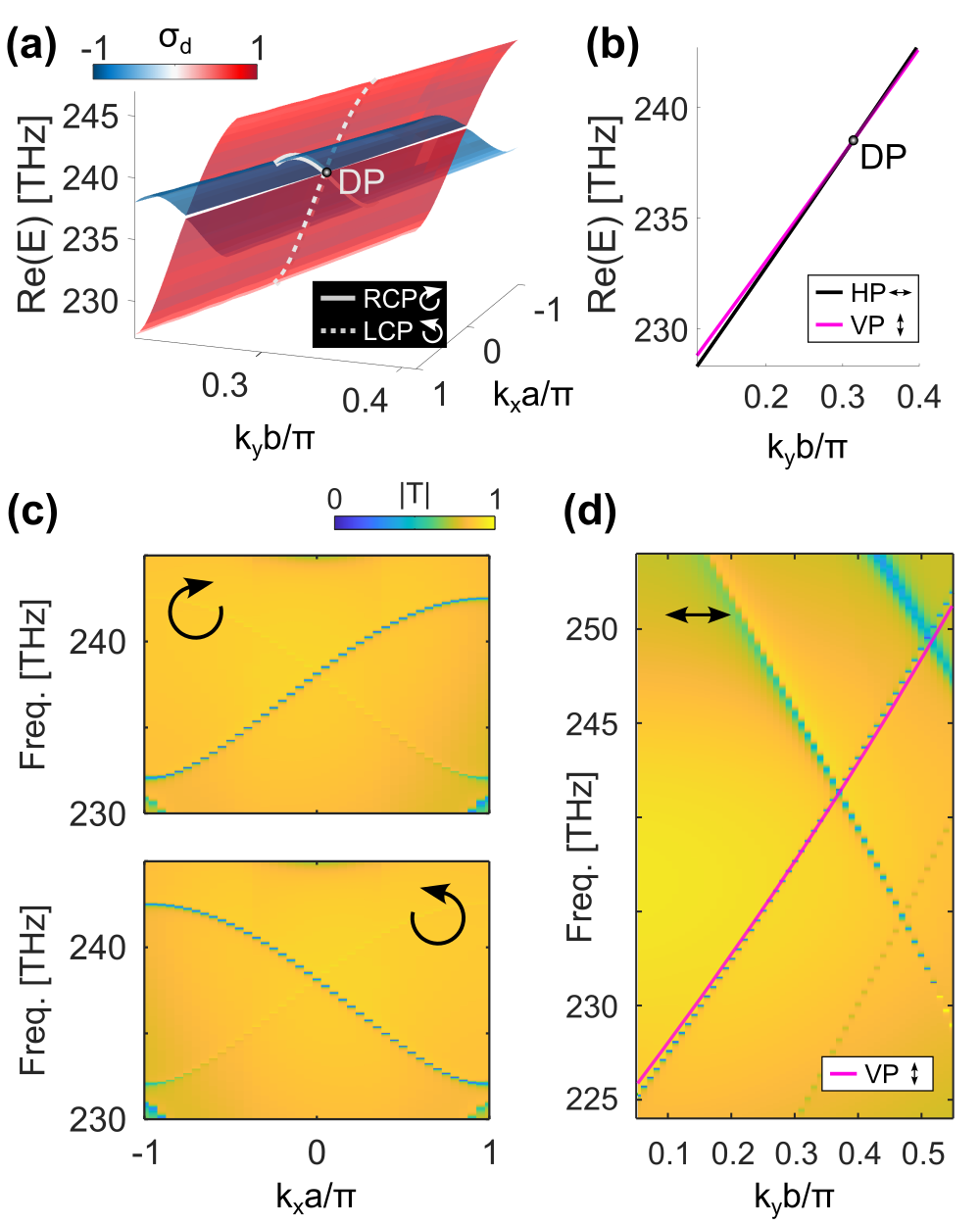}
\caption{Simulations of the metasurface. (a)~Underlying eigenmode structure: eigenenergies $E$ and color coded eigenpolarization $\sigma_d$ in 2D $k$-space. The solid and dashed white line marks the eigenenergies for a $k_x$-sweep along the Dirac point for right circular polarization (RCP) and left circular polarization (LCP), respectively. (b)~Eigenenergies for $k_x=0$ that are of linear horizontal polarization (HP, black) and vertical polarization (VP, pink).  (c)~Simulated transmission for a sweep in $k_x$ for incident RCP (top) and LCP (bottom) at $k_y=K_y=0.3\pi/b$. (d)~ Simulated transmission for a sweep in $k_y$ at $k_x=K_x=0$ for horizontally polarized incident light. The pink line marks the according resonances for vertical polarization (see Fig.~S4 in the Supporting Information).}
    \label{fig:2}
\end{figure}

%
We examine the eigenmode structure discussed above via the transmission  $|T|$. The metasurface reflects incident light that matches the polarization and frequency of the eigenmode, leading to a dip in $|T|$ at the corresponding eigenenergy (see inset graph for $|R|$ and $|T|$ in Fig.~\ref{fig:1}(a)).
The momentum-space behavior is probed by varying the angle of incidence and calculating the transmitted spectra. 
Figure \ref{fig:2}(c) depicts the simulated transmission for $k_y=0.3\pi/b=K_y$ that exhibits a resonant mode, shifting in frequency with $k_x$. 
The eigenmode, as well as the transmission simulations, were conducted using a full-wave 3D finite element method (frequency-domain) to solve the time-harmonic Maxwell equations.
For incident light of right circular polarization (RCP), the resonance frequency increases with $k_x$ forming an 
$S$-shaped curve with positive slope in the $k_x$-sweep (see Fig.~\ref{fig:2}(c), top). In contrast, the opposite/ negative slope occurs for left circular polarization (LCP), i.e.~the frequency decreases with $k_x$ (see Fig.~\ref{fig:2}(c), bottom). At $k_x=0$, the RCP and LCP modes intersect, indicating the position of the Dirac point $K_x$. 
The resonances present in the transmission spectra match the eigenmodes depicted in Figure \ref{fig:2}(a), which resembles Figure \ref{fig:1}(c) but shows absolute energies $E$ (rather than the relative energy $\pm\mathrm{Re}(E_2-E_1)/2$) and a larger region in $k$-space. The white solid line in Figure \ref{fig:2}(a) marks the eigenmodes of $\sigma_d=-1$ for a fixed value of $k_y=0.3\pi/b=K_y$, i.e.~the position of the Dirac point. The corresponding eigenenergies increase with $k_x$ and form an $S$-shaped curve like the resonances in the transmission spectra for RCP (Fig.~\ref{fig:2}(c), top). The same holds for the white dashed line that marks eigenmodes of $\sigma_d=+1$ and is in accordance with LCP transmission (Fig.~\ref{fig:2}(c), bottom).

In contrast to the circularly polarized states, the linearly polarized states are present for $k_x=0$ (see Fig.~\ref{fig:1}(d)). Figure \ref{fig:2}(b) shows a cross section of the eigenstates pictured in Figure \ref{fig:2}(a) for $k_x=0$ with color coded horizontal (HP) and vertical polarization (VP). It reveals the crossing of the two polarizations at the Dirac point and the overall positive slope of a Type-II Dirac Point~\cite{hu2018type}.
Focusing on just the upper or lower energy band, the crossing can be interpreted as a polarization swap at the Dirac point as illustrated in Figure \ref{fig:1}(d). 
The simulated transmission for linearly horizontally polarized input shown in Figure \ref{fig:2}(d) exhibits the resonance mode increasing in frequency with $k_y$ in accordance to the eigenmodes in Figure \ref{fig:2}(b). The resonances for vertically polarized input are marked by the pink line, revealing the expected crossing of these two modes at the Dirac point.

This confirms, that the eigenstate behavior in momentum-space can be studied via the transmission of differently polarized incident light.
Considering the reflected light (represented by the transmission dips), 
this metasurface provides a wave-vector depended polarization for the reflected light.
More importantly, for unpolarized illumination, the reflected polarization can be tuned to a desired state by adjusting the angle of incidence because only the polarization matching the corresponding eigenmode/eigenpolarization is reflected. This can act as a polarization filter or tunable reflective light selector.
%
%
\section{Experimental methods}  

\begin{figure}[t]
    \centering
    \includegraphics[width=\columnwidth]{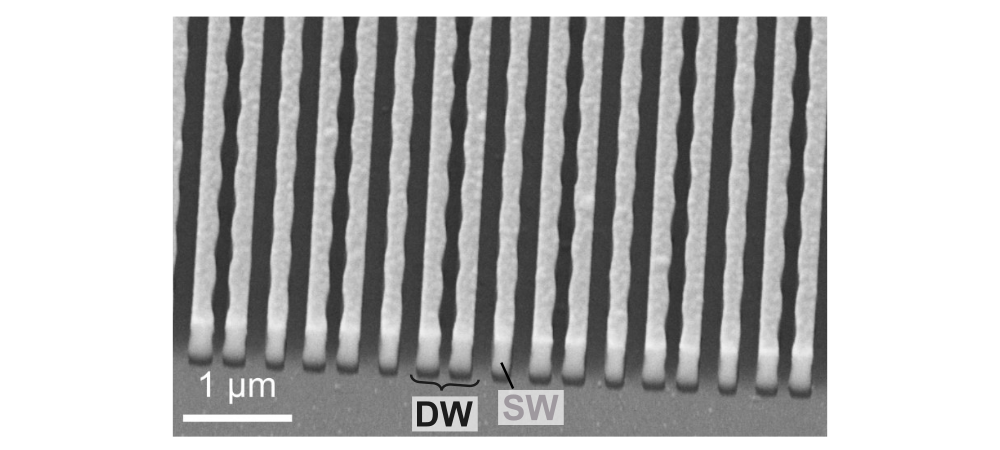}
\caption{Scanning electron micrograph of the silicon waveguide metasurface fabricated on top of a glass substrate and marked double waveguide (DW) and single waveguide (SW).}
    \label{fig:3}
\end{figure}

To investigate such metasurface experimentally, we have fabricated silicon waveguides on top of a glass substrate according to the unit cell and dimensions mentioned above by following the fabricational process steps of silicon deposition, patterning, etch-mask deposition, lift-off, etching and etch-mask removal (see Fig.~S1 in the Supporting Information).
First, amorphous silicon was deposited by plasma-enhanced chemical vapor deposition. Then, the positive tone resist Polymethylmethacrylat (PMMA) was spin coated on top and patterned by electron beam lithography. After development, \SI{15}{\nano\meter} of chromium were evaporated on top of the PMMA mask. A lift-off process reveals the etching mask for the subsequent inductively coupled plasma reactive ion etching, unveiling the silicon waveguides.
A chemical removal of the chromium mask finishes the process.  A scanning electron micrograph of the resulting structure is depicted in Figure \ref{fig:3}.

The fabricated structure is analyzed by a transmission measurement using a white light \texttt{Fianium} laser, a \texttt{Shamrock} spectrometer and polarization optics for generating and analyzing polarized light. A two-axis rotation sample holder stage is used to set $k_x$ and $k_y$. More details are given in the Supporting Information.

\section{Experimental results and discussion}  
\begin{figure}[t!]
    \centering
    \includegraphics[width=\columnwidth]{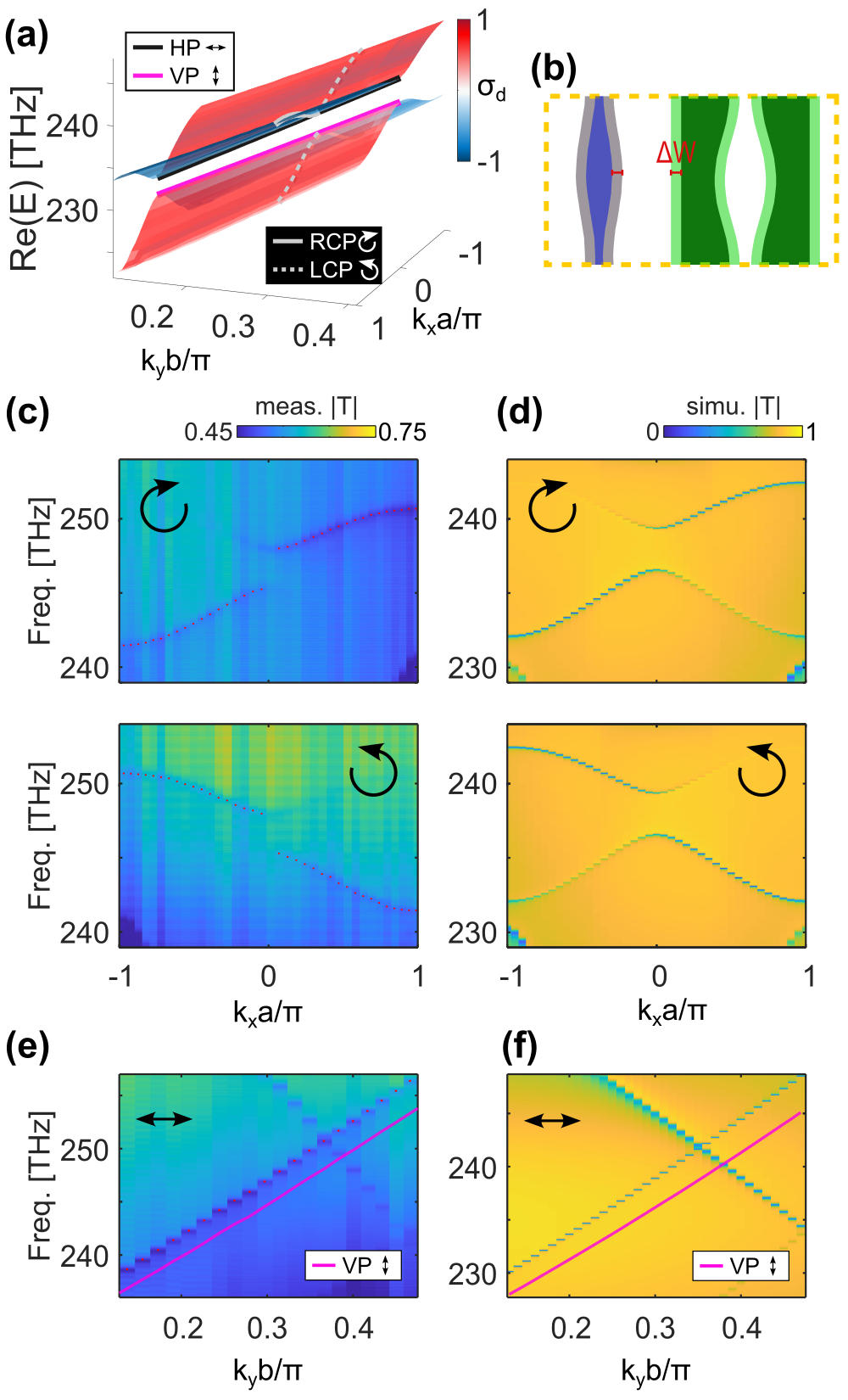}
\caption{(a)~Calculated eigenmodes of the waveguide structure with width offset $\Delta W=\SI{3.5}{\nano\meter}$. The white lines mark the relevant eigenmodes for a $k_x$-sweep for RCP (solid line) and LCP (dashed line) at $k_y=0.3\pi/b$. The black (pink) line marks the relevant eigenmodes for the $k_y$-sweep with HP (VP) at $k_x=0$. (b)~Sketch of the unit cell modified by the width offset $\Delta W$. (c)~Measured and (d)~simulated transmission for a sweep in $k_x$ at $k_y=0.3\pi/b$ for RCP (top) and LCP (bottom) input light. (e)~Measured and (f)~simulated transmission for a sweep in $k_y$ at $k_x=0$ for horizontal polarization. The pink line marks the according resonances for vertical polarization (see Fig.~S4 in the Supporting Information). The small red dots in (c) and (e) emphasize
the measured resonance position and the simulated transmission (d),~(f) apply to $\Delta W=\SI{3.5}{\nano\meter}$.}
    \label{fig:4}
\end{figure} 
Measuring the transmission of the fabricated metasurface using circular polarization results in Figure \ref{fig:4}(c). For this measurement, $k_y$ is set to the predicted Dirac point position $K_y=0.3\pi/b$ and $k_x$ is varied. For RCP, the resonance frequency increases with $k_x$ giving a positive slope (see Fig.~\ref{fig:4}(a),top) and for LCP a negative slope is observed (see Fig.~\ref{fig:4}(a),bottom) which fits the expectation/theory discussed above. Compared to the simulated transmission in Figure \ref{fig:2}(c), the $S$-shaped curve is not continuous but has a gap close to $k_x=0$ of $\Delta \mathrm{Freq}=\SI{2.7}{\tera\hertz}$ i.e.~no Dirac point is present here.
This deviation from the previously discussed behavior can be explained by a width offset $\Delta W$ that decreases the overall width of the single waveguide and increases the width of both double waveguides (see sketch in Fig.~\ref{fig:4}(b)).
With $\Delta W=\SI{3.5}{\nano\meter}$ included, the simulated transmission in Figure \ref{fig:4}(d) matches the measured data well. 
The required width offset of $\Delta W=\SI{3.5}{\nano\meter}$ corresponds to less than $\SI{2.5}{\percent}$ deviation of the total width and can be explained by the fabrication tolerance of the electron beam lithography \cite{vieu2000electron}. Figure \ref{fig:4}(a) shows the resulting eigenmodes for $\Delta W=\SI{3.5}{\nano\meter}$, indicating that the Dirac point is no longer present in the evaluated $k$-range. The white lines mark the resonance for RCP (solid line) and LCP (dashed line) at $k_y=0.3\pi/b$ emphasizing the appearance of the gap.

Comparing the measurement with the tailored simulation (Fig.~\ref{fig:4}(c) and (d) respectively), the measurement data is shifted in energy which can be explained by the deviation in the fabricated height.
Additionally, the resonances in the measurement are less dominant and the lower contrast arises mainly because of the losses in amorphous silicon. Even though the losses are low in this near-infrared regime, the imaginary refractive index of $\mathrm{Im}(n_\mathrm{Si})=0.0034$ is enough to decrease the transmission dip from about $\SI{90}{\percent}$ to $\SI{5}{\percent}$, leading to a blurred transmission image (see Figure S3 in the Supporting Information). 

Next, we consider the linearly polarized eigenstates present at $k_x=0$ that are marked by the black and pink line in Figure \ref{fig:4}(a). Since there is no Dirac point present within the pictured $k_y$-range for $\Delta W=\SI{3.5}{\nano\meter}$, the horizontally (HP) and vertically (VP) polarized modes do not cross with HP being larger in energy.
The experimental data shown in Figure \ref{fig:4}(e) reveals that the resonance for HP is increasing in frequency with $k_y$. The corresponding VP resonance frequency is marked by the pink line, confirming the absence of the Dirac point and the lower energy of the VP mode. The raw data for VP and a detailed comparison of HP and VP is given in the Supporting Information (Figure S5).
The measurement is in good agreement with the simulated transmission conducted with $\Delta W=\SI{3.5}{\nano\meter}$ shown in Figure \ref{fig:4}(f).

\begin{figure}[t!]
    \centering
    \includegraphics[width=\columnwidth]{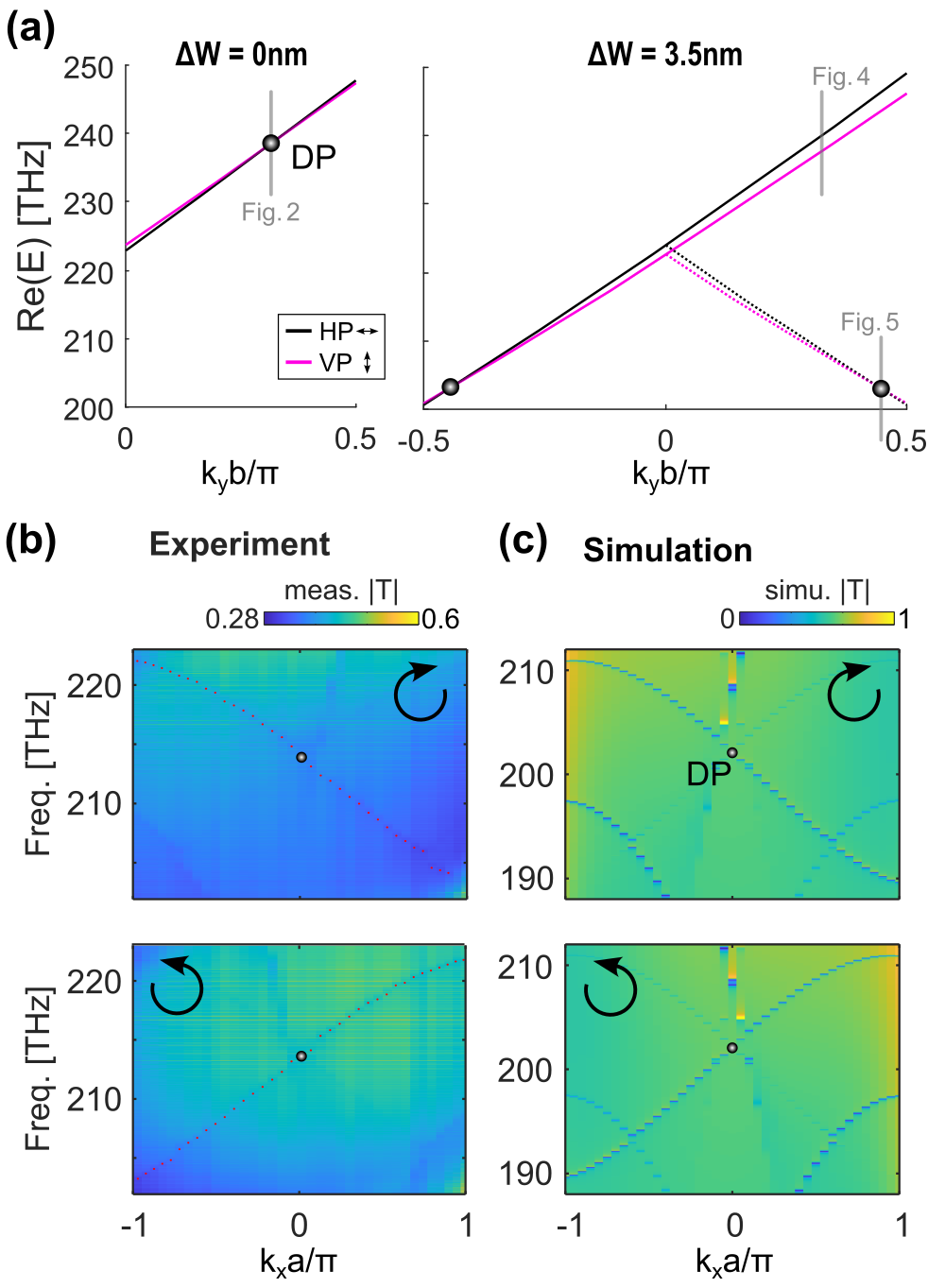}
\caption{(a)~Simulated eigenmodes along $k_y$ at $k_x=0$ without (left) and with (right) width offset $\Delta W$. The solid lines correspond to the linearly polarized eigenmodes and the dashed lines to the mirrored ones. The vertical gray lines mark the position in $k_y$ and energy of the $k_x$-sweeps discussed in the corresponding Figures. (b)~Measured and (c)~simulated transmission at $k_yb/\pi=0.46$ for RCP (top) and LCP (bottom). The red dots in (b)~mark the positions of the measured resonances. For the simulations in (c) $\Delta W=\SI{3.5}{\nano\meter}$ applies.}
    \label{fig:5}
\end{figure} 
To understand the influence of the width W, we study the effect of $\Delta W$. We found that $\Delta W$ shifts the position of the Dirac point $K_y$ (see Fig.~S7 in the Supporting Information). 
For $\Delta W=\SI{3.5}{\nano\meter}$, the Dirac point is no longer present at $K_y=0.3\pi/b$ (as in Figure \ref{fig:2}), but shifts to $K_y=-0.46\pi/b$ and $E_\mathrm{DP}=\SI{202}{\tera\hertz}$ (see Fig.~\ref{fig:5}(a)). The Dirac point position in $k_x$ remains at $K_x=0$.
To cover this shift mathematically, we extend the model derived in \cite{gao2022spin}, which is described in detail in the Supporting Information.
Because the sample rotation was limited to $k_y\geq0$ in the experiment, we investigate the mirrored Dirac point at $K_y=+0.46\pi/b$. The linearly polarized modes crossing at this mirrored Dirac point are marked by the dashed lines in Figure \ref{fig:5}(a,right) and result from a mirror inversion at $k_y=0$.
To maintain the handedness of the 3D-space consisting of the 2D $k$-space and the energy (like in Figure \ref{fig:2}(a)), the $k_x$-axis gets inverted as well or in other words: the eigenmodes at $(-k_x,-k_y)$ map to $(k_x,k_y)$ with the eigenpolarization staying unaffected.

The measured data for $k_y=K_y=+0.46\pi/b$ and circularly polarized input light are shown in Figure \ref{fig:5}(b). In comparison to the previous measurement at $k_y=0.3\pi/b$ in Figure \ref{fig:4}(c), the $S$-curve is continuous and the gap at $k_x=0$ is closed, which confirms the presence of the Dirac point and even the mirrored Dirac point.
The experimental data agrees with the simulated transmission for $\Delta W = \SI{3.5}{\nano\meter}$, shown in Figure \ref{fig:5}(c). 
The extended tight-binding model reproduces the eigenmodes in the vicinity of the Dirac point, as shown in the Supporting Information.
We experimentally observed the predicted Dirac point of the metasurface and the expected continuous $S$‑shaped resonance dispersion in $k_x$. In comparison to the theoretical expectation in Figure \ref{fig:2}(c), the slope in Figure \ref{fig:5}(b) is of opposite sign due to the inversion of the $k_x$-axis argued above.

%
\section{Conclusion} 

In conclusion, we studied the sinusoidally modulated silicon waveguide metasurface theoretically and experimentally. This structure provides circularly and linearly polarized eigenstates around the Dirac point and reflects light of the corresponding polarization and energy. Our predictions are experimentally confirmed by transmission measurements for different polarizations and angles of incidence, that can be explained well by simulations with a width offset $\Delta W=\SI{3.5}{\nano\meter}$.   
We found that the properties are very sensitive to small fabricational deviations within the electron beam lithography tolerance. A width offset of a few percentage cause the Dirac point to shift by more than \SI{200}{\percent}. 
This sensitivity, not only regarding the width, but also the height and refractive indices, makes it challenging to fabricate a metasurface that performs exactly as predicted. 
However, by determining the Dirac point position experimentally, we can achieve the desired wave-vector and polarization  dependent optical response. By adjusting the model parameters, the simulated behavior accurately reproduces the experimental observations, indicating extensive consistency between simulations and experiment.
Higher precision in fabrication could be achieved by scaling up the whole system from the nanometer to the micrometer scale, i.e.~the structure could prove valuable when working in the microwave-regime.  
In the optical regime, we have experimentally demonstrated a metasurface that utilizes near-field coupling to provide a line of circular and linear polarization in the $k$-space.
Such metasurface could allow angle-tunable polarization filters by tuning the incident $k$-vector.

\vspace{5mm}
The authors acknowledge financial support by the Deutsche Forschungsgemeinschaft (DFG, German Research Foundation) – Grant Nos. 514785315 and 467358803.
%
%

\begin{thebibliography}{28}%
\makeatletter
\providecommand \@ifxundefined [1]{%
 \@ifx{#1\undefined}
}%
\providecommand \@ifnum [1]{%
 \ifnum #1\expandafter \@firstoftwo
 \else \expandafter \@secondoftwo
 \fi
}%
\providecommand \@ifx [1]{%
 \ifx #1\expandafter \@firstoftwo
 \else \expandafter \@secondoftwo
 \fi
}%
\providecommand \natexlab [1]{#1}%
\providecommand \enquote  [1]{``#1''}%
\providecommand \bibnamefont  [1]{#1}%
\providecommand \bibfnamefont [1]{#1}%
\providecommand \citenamefont [1]{#1}%
\providecommand \href@noop [0]{\@secondoftwo}%
\providecommand \href [0]{\begingroup \@sanitize@url \@href}%
\providecommand \@href[1]{\@@startlink{#1}\@@href}%
\providecommand \@@href[1]{\endgroup#1\@@endlink}%
\providecommand \@sanitize@url [0]{\catcode `\\12\catcode `\$12\catcode
  `\&12\catcode `\#12\catcode `\^12\catcode `\_12\catcode `\%12\relax}%
\providecommand \@@startlink[1]{}%
\providecommand \@@endlink[0]{}%
\providecommand \url  [0]{\begingroup\@sanitize@url \@url }%
\providecommand \@url [1]{\endgroup\@href {#1}{\urlprefix }}%
\providecommand \urlprefix  [0]{URL }%
\providecommand \Eprint [0]{\href }%
\providecommand \doibase [0]{https://doi.org/}%
\providecommand \selectlanguage [0]{\@gobble}%
\providecommand \bibinfo  [0]{\@secondoftwo}%
\providecommand \bibfield  [0]{\@secondoftwo}%
\providecommand \translation [1]{[#1]}%
\providecommand \BibitemOpen [0]{}%
\providecommand \bibitemStop [0]{}%
\providecommand \bibitemNoStop [0]{.\EOS\space}%
\providecommand \EOS [0]{\spacefactor3000\relax}%
\providecommand \BibitemShut  [1]{\csname bibitem#1\endcsname}%
\let\auto@bib@innerbib\@empty
\bibitem [{\citenamefont {Yu}\ \emph {et~al.}(2011)\citenamefont {Yu},
  \citenamefont {Genevet}, \citenamefont {Kats}, \citenamefont {Aieta},
  \citenamefont {Tetienne}, \citenamefont {Capasso},\ and\ \citenamefont
  {Gaburro}}]{yu2011light}%
  \BibitemOpen
  \bibfield  {author} {\bibinfo {author} {\bibfnamefont {N.}~\bibnamefont
  {Yu}}, \bibinfo {author} {\bibfnamefont {P.}~\bibnamefont {Genevet}},
  \bibinfo {author} {\bibfnamefont {M.~A.}\ \bibnamefont {Kats}}, \bibinfo
  {author} {\bibfnamefont {F.}~\bibnamefont {Aieta}}, \bibinfo {author}
  {\bibfnamefont {J.-P.}\ \bibnamefont {Tetienne}}, \bibinfo {author}
  {\bibfnamefont {F.}~\bibnamefont {Capasso}},\ and\ \bibinfo {author}
  {\bibfnamefont {Z.}~\bibnamefont {Gaburro}},\ }\bibfield  {title} {\bibinfo
  {title} {Light propagation with phase discontinuities: generalized laws of
  reflection and refraction},\ }\href@noop {} {\bibfield  {journal} {\bibinfo
  {journal} {Science}\ }\textbf {\bibinfo {volume} {334}},\ \bibinfo {pages}
  {333} (\bibinfo {year} {2011})}\BibitemShut {NoStop}%
\bibitem [{\citenamefont {Spinelli}\ \emph {et~al.}(2012)\citenamefont
  {Spinelli}, \citenamefont {Verschuuren},\ and\ \citenamefont
  {Polman}}]{spinelli2012broadband}%
  \BibitemOpen
  \bibfield  {author} {\bibinfo {author} {\bibfnamefont {P.}~\bibnamefont
  {Spinelli}}, \bibinfo {author} {\bibfnamefont {M.}~\bibnamefont
  {Verschuuren}},\ and\ \bibinfo {author} {\bibfnamefont {A.}~\bibnamefont
  {Polman}},\ }\bibfield  {title} {\bibinfo {title} {Broadband omnidirectional
  antireflection coating based on subwavelength surface mie resonators},\
  }\href@noop {} {\bibfield  {journal} {\bibinfo  {journal} {Nature
  Communications}\ }\textbf {\bibinfo {volume} {3}},\ \bibinfo {pages} {692}
  (\bibinfo {year} {2012})}\BibitemShut {NoStop}%
\bibitem [{\citenamefont {Geromel}\ \emph {et~al.}(2023)\citenamefont
  {Geromel}, \citenamefont {Rennerich}, \citenamefont {Zentgraf},\ and\
  \citenamefont {Kitzerow}}]{geromel2023geometric}%
  \BibitemOpen
  \bibfield  {author} {\bibinfo {author} {\bibfnamefont {R.}~\bibnamefont
  {Geromel}}, \bibinfo {author} {\bibfnamefont {R.}~\bibnamefont {Rennerich}},
  \bibinfo {author} {\bibfnamefont {T.}~\bibnamefont {Zentgraf}},\ and\
  \bibinfo {author} {\bibfnamefont {H.}~\bibnamefont {Kitzerow}},\ }\bibfield
  {title} {\bibinfo {title} {Geometric-phase metalens to be used for tunable
  optical tweezers in microfluidics},\ }\href@noop {} {\bibfield  {journal}
  {\bibinfo  {journal} {Liquid Crystals}\ }\textbf {\bibinfo {volume} {50}},\
  \bibinfo {pages} {1193} (\bibinfo {year} {2023})}\BibitemShut {NoStop}%
\bibitem [{\citenamefont {Zheng}\ \emph {et~al.}(2015)\citenamefont {Zheng},
  \citenamefont {M{\"u}hlenbernd}, \citenamefont {Kenney}, \citenamefont {Li},
  \citenamefont {Zentgraf},\ and\ \citenamefont
  {Zhang}}]{zheng2015metasurface}%
  \BibitemOpen
  \bibfield  {author} {\bibinfo {author} {\bibfnamefont {G.}~\bibnamefont
  {Zheng}}, \bibinfo {author} {\bibfnamefont {H.}~\bibnamefont
  {M{\"u}hlenbernd}}, \bibinfo {author} {\bibfnamefont {M.}~\bibnamefont
  {Kenney}}, \bibinfo {author} {\bibfnamefont {G.}~\bibnamefont {Li}}, \bibinfo
  {author} {\bibfnamefont {T.}~\bibnamefont {Zentgraf}},\ and\ \bibinfo
  {author} {\bibfnamefont {S.}~\bibnamefont {Zhang}},\ }\bibfield  {title}
  {\bibinfo {title} {Metasurface holograms reaching 80\% efficiency},\
  }\href@noop {} {\bibfield  {journal} {\bibinfo  {journal} {Nature
  Nanotechnology}\ }\textbf {\bibinfo {volume} {10}},\ \bibinfo {pages} {308}
  (\bibinfo {year} {2015})}\BibitemShut {NoStop}%
\bibitem [{\citenamefont {Huang}\ \emph {et~al.}(2013)\citenamefont {Huang},
  \citenamefont {Chen}, \citenamefont {M{\"u}hlenbernd}, \citenamefont {Zhang},
  \citenamefont {Chen}, \citenamefont {Bai}, \citenamefont {Tan}, \citenamefont
  {Jin}, \citenamefont {Cheah}, \citenamefont {Qiu} \emph
  {et~al.}}]{huang2013three}%
  \BibitemOpen
  \bibfield  {author} {\bibinfo {author} {\bibfnamefont {L.}~\bibnamefont
  {Huang}}, \bibinfo {author} {\bibfnamefont {X.}~\bibnamefont {Chen}},
  \bibinfo {author} {\bibfnamefont {H.}~\bibnamefont {M{\"u}hlenbernd}},
  \bibinfo {author} {\bibfnamefont {H.}~\bibnamefont {Zhang}}, \bibinfo
  {author} {\bibfnamefont {S.}~\bibnamefont {Chen}}, \bibinfo {author}
  {\bibfnamefont {B.}~\bibnamefont {Bai}}, \bibinfo {author} {\bibfnamefont
  {Q.}~\bibnamefont {Tan}}, \bibinfo {author} {\bibfnamefont {G.}~\bibnamefont
  {Jin}}, \bibinfo {author} {\bibfnamefont {K.-W.}\ \bibnamefont {Cheah}},
  \bibinfo {author} {\bibfnamefont {C.-W.}\ \bibnamefont {Qiu}}, \emph
  {et~al.},\ }\bibfield  {title} {\bibinfo {title} {Three-dimensional optical
  holography using a plasmonic metasurface},\ }\href@noop {} {\bibfield
  {journal} {\bibinfo  {journal} {Nature Communications}\ }\textbf {\bibinfo
  {volume} {4}},\ \bibinfo {pages} {2808} (\bibinfo {year} {2013})}\BibitemShut
  {NoStop}%
\bibitem [{\citenamefont {Klein}\ \emph {et~al.}(2006)\citenamefont {Klein},
  \citenamefont {Enkrich}, \citenamefont {Wegener},\ and\ \citenamefont
  {Linden}}]{klein2006second}%
  \BibitemOpen
  \bibfield  {author} {\bibinfo {author} {\bibfnamefont {M.~W.}\ \bibnamefont
  {Klein}}, \bibinfo {author} {\bibfnamefont {C.}~\bibnamefont {Enkrich}},
  \bibinfo {author} {\bibfnamefont {M.}~\bibnamefont {Wegener}},\ and\ \bibinfo
  {author} {\bibfnamefont {S.}~\bibnamefont {Linden}},\ }\bibfield  {title}
  {\bibinfo {title} {Second-harmonic generation from magnetic metamaterials},\
  }\href@noop {} {\bibfield  {journal} {\bibinfo  {journal} {Science}\ }\textbf
  {\bibinfo {volume} {313}},\ \bibinfo {pages} {502} (\bibinfo {year}
  {2006})}\BibitemShut {NoStop}%
\bibitem [{\citenamefont {Zhang}\ \emph {et~al.}(2013)\citenamefont {Zhang},
  \citenamefont {Wen}, \citenamefont {Zhen}, \citenamefont {Nordlander},\ and\
  \citenamefont {Halas}}]{zhang2013coherent}%
  \BibitemOpen
  \bibfield  {author} {\bibinfo {author} {\bibfnamefont {Y.}~\bibnamefont
  {Zhang}}, \bibinfo {author} {\bibfnamefont {F.}~\bibnamefont {Wen}}, \bibinfo
  {author} {\bibfnamefont {Y.-R.}\ \bibnamefont {Zhen}}, \bibinfo {author}
  {\bibfnamefont {P.}~\bibnamefont {Nordlander}},\ and\ \bibinfo {author}
  {\bibfnamefont {N.~J.}\ \bibnamefont {Halas}},\ }\bibfield  {title} {\bibinfo
  {title} {Coherent fano resonances in a plasmonic nanocluster enhance optical
  four-wave mixing},\ }\href@noop {} {\bibfield  {journal} {\bibinfo  {journal}
  {Proceedings of the National Academy of Sciences}\ }\textbf {\bibinfo
  {volume} {110}},\ \bibinfo {pages} {9215} (\bibinfo {year}
  {2013})}\BibitemShut {NoStop}%
\bibitem [{\citenamefont {Li}\ \emph {et~al.}(2017)\citenamefont {Li},
  \citenamefont {Zhang},\ and\ \citenamefont {Zentgraf}}]{li2017nonlinear}%
  \BibitemOpen
  \bibfield  {author} {\bibinfo {author} {\bibfnamefont {G.}~\bibnamefont
  {Li}}, \bibinfo {author} {\bibfnamefont {S.}~\bibnamefont {Zhang}},\ and\
  \bibinfo {author} {\bibfnamefont {T.}~\bibnamefont {Zentgraf}},\ }\bibfield
  {title} {\bibinfo {title} {Nonlinear photonic metasurfaces},\ }\href@noop {}
  {\bibfield  {journal} {\bibinfo  {journal} {Nature Reviews Materials}\
  }\textbf {\bibinfo {volume} {2}},\ \bibinfo {pages} {1} (\bibinfo {year}
  {2017})}\BibitemShut {NoStop}%
\bibitem [{\citenamefont {Meng}\ \emph {et~al.}(2021)\citenamefont {Meng},
  \citenamefont {Chen}, \citenamefont {Lu}, \citenamefont {Ding}, \citenamefont
  {Cusano}, \citenamefont {Fan}, \citenamefont {Hu}, \citenamefont {Wang},
  \citenamefont {Xie}, \citenamefont {Liu} \emph {et~al.}}]{meng2021optical}%
  \BibitemOpen
  \bibfield  {author} {\bibinfo {author} {\bibfnamefont {Y.}~\bibnamefont
  {Meng}}, \bibinfo {author} {\bibfnamefont {Y.}~\bibnamefont {Chen}}, \bibinfo
  {author} {\bibfnamefont {L.}~\bibnamefont {Lu}}, \bibinfo {author}
  {\bibfnamefont {Y.}~\bibnamefont {Ding}}, \bibinfo {author} {\bibfnamefont
  {A.}~\bibnamefont {Cusano}}, \bibinfo {author} {\bibfnamefont {J.~A.}\
  \bibnamefont {Fan}}, \bibinfo {author} {\bibfnamefont {Q.}~\bibnamefont
  {Hu}}, \bibinfo {author} {\bibfnamefont {K.}~\bibnamefont {Wang}}, \bibinfo
  {author} {\bibfnamefont {Z.}~\bibnamefont {Xie}}, \bibinfo {author}
  {\bibfnamefont {Z.}~\bibnamefont {Liu}}, \emph {et~al.},\ }\bibfield  {title}
  {\bibinfo {title} {Optical meta-waveguides for integrated photonics and
  beyond},\ }\href@noop {} {\bibfield  {journal} {\bibinfo  {journal} {Light:
  Science \& Applications}\ }\textbf {\bibinfo {volume} {10}},\ \bibinfo
  {pages} {235} (\bibinfo {year} {2021})}\BibitemShut {NoStop}%
\bibitem [{\citenamefont {Ou}\ \emph {et~al.}(2023)\citenamefont {Ou},
  \citenamefont {Wan}, \citenamefont {Wang}, \citenamefont {Zhu}, \citenamefont
  {Dong}, \citenamefont {He}, \citenamefont {Yang}, \citenamefont {Wei},
  \citenamefont {Wang},\ and\ \citenamefont {Cheng}}]{ou2023advances}%
  \BibitemOpen
  \bibfield  {author} {\bibinfo {author} {\bibfnamefont {K.}~\bibnamefont
  {Ou}}, \bibinfo {author} {\bibfnamefont {H.}~\bibnamefont {Wan}}, \bibinfo
  {author} {\bibfnamefont {G.}~\bibnamefont {Wang}}, \bibinfo {author}
  {\bibfnamefont {J.}~\bibnamefont {Zhu}}, \bibinfo {author} {\bibfnamefont
  {S.}~\bibnamefont {Dong}}, \bibinfo {author} {\bibfnamefont {T.}~\bibnamefont
  {He}}, \bibinfo {author} {\bibfnamefont {H.}~\bibnamefont {Yang}}, \bibinfo
  {author} {\bibfnamefont {Z.}~\bibnamefont {Wei}}, \bibinfo {author}
  {\bibfnamefont {Z.}~\bibnamefont {Wang}},\ and\ \bibinfo {author}
  {\bibfnamefont {X.}~\bibnamefont {Cheng}},\ }\bibfield  {title} {\bibinfo
  {title} {Advances in meta-optics and metasurfaces: Fundamentals and
  applications},\ }\href@noop {} {\bibfield  {journal} {\bibinfo  {journal}
  {Nanomaterials}\ }\textbf {\bibinfo {volume} {13}},\ \bibinfo {pages} {1235}
  (\bibinfo {year} {2023})}\BibitemShut {NoStop}%
\bibitem [{\citenamefont {Georgi}\ \emph {et~al.}(2019)\citenamefont {Georgi},
  \citenamefont {Massaro}, \citenamefont {Luo}, \citenamefont {Sain},
  \citenamefont {Montaut}, \citenamefont {Herrmann}, \citenamefont {Weiss},
  \citenamefont {Li}, \citenamefont {Silberhorn},\ and\ \citenamefont
  {Zentgraf}}]{georgi2019metasurface}%
  \BibitemOpen
  \bibfield  {author} {\bibinfo {author} {\bibfnamefont {P.}~\bibnamefont
  {Georgi}}, \bibinfo {author} {\bibfnamefont {M.}~\bibnamefont {Massaro}},
  \bibinfo {author} {\bibfnamefont {K.-H.}\ \bibnamefont {Luo}}, \bibinfo
  {author} {\bibfnamefont {B.}~\bibnamefont {Sain}}, \bibinfo {author}
  {\bibfnamefont {N.}~\bibnamefont {Montaut}}, \bibinfo {author} {\bibfnamefont
  {H.}~\bibnamefont {Herrmann}}, \bibinfo {author} {\bibfnamefont
  {T.}~\bibnamefont {Weiss}}, \bibinfo {author} {\bibfnamefont
  {G.}~\bibnamefont {Li}}, \bibinfo {author} {\bibfnamefont {C.}~\bibnamefont
  {Silberhorn}},\ and\ \bibinfo {author} {\bibfnamefont {T.}~\bibnamefont
  {Zentgraf}},\ }\bibfield  {title} {\bibinfo {title} {Metasurface
  interferometry toward quantum sensors},\ }\href@noop {} {\bibfield  {journal}
  {\bibinfo  {journal} {Light: Science \& Applications}\ }\textbf {\bibinfo
  {volume} {8}},\ \bibinfo {pages} {70} (\bibinfo {year} {2019})}\BibitemShut
  {NoStop}%
\bibitem [{\citenamefont {Kim}\ \emph {et~al.}(2021)\citenamefont {Kim},
  \citenamefont {Rana}, \citenamefont {Kim}, \citenamefont {Kim}, \citenamefont
  {Badloe}, \citenamefont {Zubair}, \citenamefont {Mehmood},\ and\
  \citenamefont {Rho}}]{kim_chiroptical_2021}%
  \BibitemOpen
  \bibfield  {author} {\bibinfo {author} {\bibfnamefont {J.}~\bibnamefont
  {Kim}}, \bibinfo {author} {\bibfnamefont {A.~S.}\ \bibnamefont {Rana}},
  \bibinfo {author} {\bibfnamefont {Y.}~\bibnamefont {Kim}}, \bibinfo {author}
  {\bibfnamefont {I.}~\bibnamefont {Kim}}, \bibinfo {author} {\bibfnamefont
  {T.}~\bibnamefont {Badloe}}, \bibinfo {author} {\bibfnamefont
  {M.}~\bibnamefont {Zubair}}, \bibinfo {author} {\bibfnamefont {M.~Q.}\
  \bibnamefont {Mehmood}},\ and\ \bibinfo {author} {\bibfnamefont
  {J.}~\bibnamefont {Rho}},\ }\bibfield  {title} {{\bibinfo
  {title} {Chiroptical {Metasurfaces}: {Principles}, {Classification}, and
  {Applications}}},\ }\href@noop {} {\bibfield  {journal} {\bibinfo  {journal}
  {Sensors}\ }\textbf {\bibinfo {volume} {21}},\ \bibinfo {pages} {4381}
  (\bibinfo {year} {2021})}\BibitemShut {NoStop}%
\bibitem [{\citenamefont {Sch{\"a}ferling}\ \emph {et~al.}(2012)\citenamefont
  {Sch{\"a}ferling}, \citenamefont {Dregely}, \citenamefont {Hentschel},\ and\
  \citenamefont {Giessen}}]{schaferling2012tailoring}%
  \BibitemOpen
  \bibfield  {author} {\bibinfo {author} {\bibfnamefont {M.}~\bibnamefont
  {Sch{\"a}ferling}}, \bibinfo {author} {\bibfnamefont {D.}~\bibnamefont
  {Dregely}}, \bibinfo {author} {\bibfnamefont {M.}~\bibnamefont {Hentschel}},\
  and\ \bibinfo {author} {\bibfnamefont {H.}~\bibnamefont {Giessen}},\
  }\bibfield  {title} {\bibinfo {title} {Tailoring enhanced optical chirality:
  design principles for chiral plasmonic nanostructures},\ }\href@noop {}
  {\bibfield  {journal} {\bibinfo  {journal} {Physical Review X}\ }\textbf
  {\bibinfo {volume} {2}},\ \bibinfo {pages} {031010} (\bibinfo {year}
  {2012})}\BibitemShut {NoStop}%
\bibitem [{\citenamefont {Mandal}(2018)}]{mandal2018large}%
  \BibitemOpen
  \bibfield  {author} {\bibinfo {author} {\bibfnamefont {P.}~\bibnamefont
  {Mandal}},\ }\bibfield  {title} {\bibinfo {title} {Large circular dichroism
  in mdm plasmonic metasurface with subwavelength crescent aperture},\
  }\href@noop {} {\bibfield  {journal} {\bibinfo  {journal} {Plasmonics}\
  }\textbf {\bibinfo {volume} {13}},\ \bibinfo {pages} {2229} (\bibinfo {year}
  {2018})}\BibitemShut {NoStop}%
\bibitem [{\citenamefont {Zhu}\ \emph {et~al.}(2018)\citenamefont {Zhu},
  \citenamefont {Chen}, \citenamefont {Zaidi}, \citenamefont {Huang},
  \citenamefont {Khorasaninejad}, \citenamefont {Sanjeev}, \citenamefont
  {Qiu},\ and\ \citenamefont {Capasso}}]{zhu2018giant}%
  \BibitemOpen
  \bibfield  {author} {\bibinfo {author} {\bibfnamefont {A.~Y.}\ \bibnamefont
  {Zhu}}, \bibinfo {author} {\bibfnamefont {W.~T.}\ \bibnamefont {Chen}},
  \bibinfo {author} {\bibfnamefont {A.}~\bibnamefont {Zaidi}}, \bibinfo
  {author} {\bibfnamefont {Y.-W.}\ \bibnamefont {Huang}}, \bibinfo {author}
  {\bibfnamefont {M.}~\bibnamefont {Khorasaninejad}}, \bibinfo {author}
  {\bibfnamefont {V.}~\bibnamefont {Sanjeev}}, \bibinfo {author} {\bibfnamefont
  {C.-W.}\ \bibnamefont {Qiu}},\ and\ \bibinfo {author} {\bibfnamefont
  {F.}~\bibnamefont {Capasso}},\ }\bibfield  {title} {\bibinfo {title} {Giant
  intrinsic chiro-optical activity in planar dielectric nanostructures},\
  }\href@noop {} {\bibfield  {journal} {\bibinfo  {journal} {Light: Science \&
  Applications}\ }\textbf {\bibinfo {volume} {7}},\ \bibinfo {pages} {17158}
  (\bibinfo {year} {2018})}\BibitemShut {NoStop}%
\bibitem [{\citenamefont {Hu}\ \emph {et~al.}(2017)\citenamefont {Hu},
  \citenamefont {Zhao}, \citenamefont {Lin}, \citenamefont {Zhu}, \citenamefont
  {Zhu}, \citenamefont {Guo}, \citenamefont {Cao},\ and\ \citenamefont
  {Wang}}]{hu2017all}%
  \BibitemOpen
  \bibfield  {author} {\bibinfo {author} {\bibfnamefont {J.}~\bibnamefont
  {Hu}}, \bibinfo {author} {\bibfnamefont {X.}~\bibnamefont {Zhao}}, \bibinfo
  {author} {\bibfnamefont {Y.}~\bibnamefont {Lin}}, \bibinfo {author}
  {\bibfnamefont {A.}~\bibnamefont {Zhu}}, \bibinfo {author} {\bibfnamefont
  {X.}~\bibnamefont {Zhu}}, \bibinfo {author} {\bibfnamefont {P.}~\bibnamefont
  {Guo}}, \bibinfo {author} {\bibfnamefont {B.}~\bibnamefont {Cao}},\ and\
  \bibinfo {author} {\bibfnamefont {C.}~\bibnamefont {Wang}},\ }\bibfield
  {title} {\bibinfo {title} {All-dielectric metasurface circular dichroism
  waveplate},\ }\href@noop {} {\bibfield  {journal} {\bibinfo  {journal}
  {Scientific Reports}\ }\textbf {\bibinfo {volume} {7}},\ \bibinfo {pages}
  {41893} (\bibinfo {year} {2017})}\BibitemShut {NoStop}%
\bibitem [{\citenamefont {Khaliq}\ \emph {et~al.}(2023)\citenamefont {Khaliq},
  \citenamefont {Nauman}, \citenamefont {Lee},\ and\ \citenamefont
  {Kim}}]{khaliq_recent_2023}%
  \BibitemOpen
  \bibfield  {author} {\bibinfo {author} {\bibfnamefont {H.~S.}\ \bibnamefont
  {Khaliq}}, \bibinfo {author} {\bibfnamefont {A.}~\bibnamefont {Nauman}},
  \bibinfo {author} {\bibfnamefont {J.}~\bibnamefont {Lee}},\ and\ \bibinfo
  {author} {\bibfnamefont {H.}~\bibnamefont {Kim}},\ }\bibfield  {title}
  {{\bibinfo {title} {Recent {Progress} on {Plasmonic} and
  {Dielectric} {Chiral} {Metasurfaces}: {Fundamentals}, {Design} {Strategies},
  and {Implementation}}},\ }\href@noop {} {\bibfield  {journal} {\bibinfo
  {journal} {Advanced Optical Materials}\ }\textbf {\bibinfo {volume} {11}},\
  \bibinfo {pages} {2300644} (\bibinfo {year} {2023})}\BibitemShut {NoStop}%
\bibitem [{\citenamefont {Shi}\ \emph {et~al.}(2022)\citenamefont {Shi},
  \citenamefont {Deng}, \citenamefont {Geng}, \citenamefont {Zeng},
  \citenamefont {Zeng}, \citenamefont {Hu}, \citenamefont {Overvig},
  \citenamefont {Li}, \citenamefont {Qiu}, \citenamefont {Alù}, \citenamefont
  {Kivshar},\ and\ \citenamefont {Li}}]{shi_planar_2022}%
  \BibitemOpen
  \bibfield  {author} {\bibinfo {author} {\bibfnamefont {T.}~\bibnamefont
  {Shi}}, \bibinfo {author} {\bibfnamefont {Z.-L.}\ \bibnamefont {Deng}},
  \bibinfo {author} {\bibfnamefont {G.}~\bibnamefont {Geng}}, \bibinfo {author}
  {\bibfnamefont {X.}~\bibnamefont {Zeng}}, \bibinfo {author} {\bibfnamefont
  {Y.}~\bibnamefont {Zeng}}, \bibinfo {author} {\bibfnamefont {G.}~\bibnamefont
  {Hu}}, \bibinfo {author} {\bibfnamefont {A.}~\bibnamefont {Overvig}},
  \bibinfo {author} {\bibfnamefont {J.}~\bibnamefont {Li}}, \bibinfo {author}
  {\bibfnamefont {C.-W.}\ \bibnamefont {Qiu}}, \bibinfo {author} {\bibfnamefont
  {A.}~\bibnamefont {Alù}}, \bibinfo {author} {\bibfnamefont {Y.~S.}\
  \bibnamefont {Kivshar}},\ and\ \bibinfo {author} {\bibfnamefont
  {X.}~\bibnamefont {Li}},\ }\bibfield  {title} {{\bibinfo
  {title} {Planar chiral metasurfaces with maximal and tunable chiroptical
  response driven by bound states in the continuum}},\ }\href@noop {}
  {\bibfield  {journal} {\bibinfo  {journal} {Nature Communications}\ }\textbf
  {\bibinfo {volume} {13}},\ \bibinfo {pages} {4111} (\bibinfo {year}
  {2022})}\BibitemShut {NoStop}%
\bibitem [{\citenamefont {Zentgraf}\ \emph {et~al.}(2004)\citenamefont
  {Zentgraf}, \citenamefont {Christ}, \citenamefont {Kuhl},\ and\ \citenamefont
  {Giessen}}]{zentgraf2004tailoring}%
  \BibitemOpen
  \bibfield  {author} {\bibinfo {author} {\bibfnamefont {T.}~\bibnamefont
  {Zentgraf}}, \bibinfo {author} {\bibfnamefont {A.}~\bibnamefont {Christ}},
  \bibinfo {author} {\bibfnamefont {J.}~\bibnamefont {Kuhl}},\ and\ \bibinfo
  {author} {\bibfnamefont {H.}~\bibnamefont {Giessen}},\ }\bibfield  {title}
  {\bibinfo {title} {Tailoring the ultrafast dephasing of quasiparticles in
  metallic photonic crystals},\ }\href@noop {} {\bibfield  {journal} {\bibinfo
  {journal} {Physical Review Letters}\ }\textbf {\bibinfo {volume} {93}},\
  \bibinfo {pages} {243901} (\bibinfo {year} {2004})}\BibitemShut {NoStop}%
\bibitem [{\citenamefont {Kodigala}\ \emph {et~al.}(2017)\citenamefont
  {Kodigala}, \citenamefont {Lepetit}, \citenamefont {Gu}, \citenamefont
  {Bahari}, \citenamefont {Fainman},\ and\ \citenamefont
  {Kant{\'e}}}]{kodigala2017lasing}%
  \BibitemOpen
  \bibfield  {author} {\bibinfo {author} {\bibfnamefont {A.}~\bibnamefont
  {Kodigala}}, \bibinfo {author} {\bibfnamefont {T.}~\bibnamefont {Lepetit}},
  \bibinfo {author} {\bibfnamefont {Q.}~\bibnamefont {Gu}}, \bibinfo {author}
  {\bibfnamefont {B.}~\bibnamefont {Bahari}}, \bibinfo {author} {\bibfnamefont
  {Y.}~\bibnamefont {Fainman}},\ and\ \bibinfo {author} {\bibfnamefont
  {B.}~\bibnamefont {Kant{\'e}}},\ }\bibfield  {title} {\bibinfo {title}
  {Lasing action from photonic bound states in continuum},\ }\href@noop {}
  {\bibfield  {journal} {\bibinfo  {journal} {Nature}\ }\textbf {\bibinfo
  {volume} {541}},\ \bibinfo {pages} {196} (\bibinfo {year}
  {2017})}\BibitemShut {NoStop}%
\bibitem [{\citenamefont {Koshelev}\ \emph {et~al.}(2020)\citenamefont
  {Koshelev}, \citenamefont {Kruk}, \citenamefont {Melik-Gaykazyan},
  \citenamefont {Choi}, \citenamefont {Bogdanov}, \citenamefont {Park},\ and\
  \citenamefont {Kivshar}}]{koshelev2020subwavelength}%
  \BibitemOpen
  \bibfield  {author} {\bibinfo {author} {\bibfnamefont {K.}~\bibnamefont
  {Koshelev}}, \bibinfo {author} {\bibfnamefont {S.}~\bibnamefont {Kruk}},
  \bibinfo {author} {\bibfnamefont {E.}~\bibnamefont {Melik-Gaykazyan}},
  \bibinfo {author} {\bibfnamefont {J.-H.}\ \bibnamefont {Choi}}, \bibinfo
  {author} {\bibfnamefont {A.}~\bibnamefont {Bogdanov}}, \bibinfo {author}
  {\bibfnamefont {H.-G.}\ \bibnamefont {Park}},\ and\ \bibinfo {author}
  {\bibfnamefont {Y.}~\bibnamefont {Kivshar}},\ }\bibfield  {title} {\bibinfo
  {title} {Subwavelength dielectric resonators for nonlinear nanophotonics},\
  }\href@noop {} {\bibfield  {journal} {\bibinfo  {journal} {Science}\ }\textbf
  {\bibinfo {volume} {367}},\ \bibinfo {pages} {288} (\bibinfo {year}
  {2020})}\BibitemShut {NoStop}%
\bibitem [{\citenamefont {Christ}\ \emph {et~al.}(2006)\citenamefont {Christ},
  \citenamefont {Zentgraf}, \citenamefont {Tikhodeev}, \citenamefont {Gippius},
  \citenamefont {Kuhl},\ and\ \citenamefont {Giessen}}]{christ2006controlling}%
  \BibitemOpen
  \bibfield  {author} {\bibinfo {author} {\bibfnamefont {A.}~\bibnamefont
  {Christ}}, \bibinfo {author} {\bibfnamefont {T.}~\bibnamefont {Zentgraf}},
  \bibinfo {author} {\bibfnamefont {S.}~\bibnamefont {Tikhodeev}}, \bibinfo
  {author} {\bibfnamefont {N.}~\bibnamefont {Gippius}}, \bibinfo {author}
  {\bibfnamefont {J.}~\bibnamefont {Kuhl}},\ and\ \bibinfo {author}
  {\bibfnamefont {H.}~\bibnamefont {Giessen}},\ }\bibfield  {title} {\bibinfo
  {title} {Controlling the interaction between localized and delocalized
  surface plasmon modes: Experiment and numerical calculations},\ }\href@noop
  {} {\bibfield  {journal} {\bibinfo  {journal} {Physical Review B—Condensed
  Matter and Materials Physics}\ }\textbf {\bibinfo {volume} {74}},\ \bibinfo
  {pages} {155435} (\bibinfo {year} {2006})}\BibitemShut {NoStop}%
\bibitem [{\citenamefont {Liu}\ \emph {et~al.}(2019)\citenamefont {Liu},
  \citenamefont {Wang}, \citenamefont {Zhang}, \citenamefont {Wang},
  \citenamefont {Zhao}, \citenamefont {Guan}, \citenamefont {Liu},
  \citenamefont {Shi},\ and\ \citenamefont {Zi}}]{liu_circularly_2019}%
  \BibitemOpen
  \bibfield  {author} {\bibinfo {author} {\bibfnamefont {W.}~\bibnamefont
  {Liu}}, \bibinfo {author} {\bibfnamefont {B.}~\bibnamefont {Wang}}, \bibinfo
  {author} {\bibfnamefont {Y.}~\bibnamefont {Zhang}}, \bibinfo {author}
  {\bibfnamefont {J.}~\bibnamefont {Wang}}, \bibinfo {author} {\bibfnamefont
  {M.}~\bibnamefont {Zhao}}, \bibinfo {author} {\bibfnamefont {F.}~\bibnamefont
  {Guan}}, \bibinfo {author} {\bibfnamefont {X.}~\bibnamefont {Liu}}, \bibinfo
  {author} {\bibfnamefont {L.}~\bibnamefont {Shi}},\ and\ \bibinfo {author}
  {\bibfnamefont {J.}~\bibnamefont {Zi}},\ }\bibfield  {title}
  {{\bibinfo {title} {Circularly {Polarized} {States}
  {Spawning} from {Bound} {States} in the {Continuum}}},\ }\href@noop {}
  {\bibfield  {journal} {\bibinfo  {journal} {Physical Review Letters}\
  }\textbf {\bibinfo {volume} {123}},\ \bibinfo {pages} {116104} (\bibinfo
  {year} {2019})}\BibitemShut {NoStop}%
\bibitem [{\citenamefont {Gao}\ \emph {et~al.}(2022)\citenamefont {Gao},
  \citenamefont {Sain},\ and\ \citenamefont {Zentgraf}}]{gao2022spin}%
  \BibitemOpen
  \bibfield  {author} {\bibinfo {author} {\bibfnamefont {W.}~\bibnamefont
  {Gao}}, \bibinfo {author} {\bibfnamefont {B.}~\bibnamefont {Sain}},\ and\
  \bibinfo {author} {\bibfnamefont {T.}~\bibnamefont {Zentgraf}},\ }\bibfield
  {title} {\bibinfo {title} {Spin-orbit interaction of light enabled by
  negative coupling in high-quality-factor optical metasurfaces},\ }\href@noop
  {} {\bibfield  {journal} {\bibinfo  {journal} {Physical Review Applied}\
  }\textbf {\bibinfo {volume} {17}},\ \bibinfo {pages} {044022} (\bibinfo
  {year} {2022})}\BibitemShut {NoStop}%
\bibitem [{\citenamefont {Barton~III}\ \emph {et~al.}(2021)\citenamefont
  {Barton~III}, \citenamefont {Lawrence},\ and\ \citenamefont
  {Dionne}}]{barton2021wavefront}%
  \BibitemOpen
  \bibfield  {author} {\bibinfo {author} {\bibfnamefont {D.}~\bibnamefont
  {Barton~III}}, \bibinfo {author} {\bibfnamefont {M.}~\bibnamefont
  {Lawrence}},\ and\ \bibinfo {author} {\bibfnamefont {J.}~\bibnamefont
  {Dionne}},\ }\bibfield  {title} {\bibinfo {title} {Wavefront shaping and
  modulation with resonant electro-optic phase gradient metasurfaces},\
  }\href@noop {} {\bibfield  {journal} {\bibinfo  {journal} {Applied Physics
  Letters}\ }\textbf {\bibinfo {volume} {118}} (\bibinfo {year}
  {2021})}\BibitemShut {NoStop}%
\bibitem [{\citenamefont {Lawrence}\ \emph {et~al.}(2020)\citenamefont
  {Lawrence}, \citenamefont {Barton~III}, \citenamefont {Dixon}, \citenamefont
  {Song}, \citenamefont {van~de Groep}, \citenamefont {Brongersma},\ and\
  \citenamefont {Dionne}}]{lawrence2020high}%
  \BibitemOpen
  \bibfield  {author} {\bibinfo {author} {\bibfnamefont {M.}~\bibnamefont
  {Lawrence}}, \bibinfo {author} {\bibfnamefont {D.~R.}\ \bibnamefont
  {Barton~III}}, \bibinfo {author} {\bibfnamefont {J.}~\bibnamefont {Dixon}},
  \bibinfo {author} {\bibfnamefont {J.-H.}\ \bibnamefont {Song}}, \bibinfo
  {author} {\bibfnamefont {J.}~\bibnamefont {van~de Groep}}, \bibinfo {author}
  {\bibfnamefont {M.~L.}\ \bibnamefont {Brongersma}},\ and\ \bibinfo {author}
  {\bibfnamefont {J.~A.}\ \bibnamefont {Dionne}},\ }\bibfield  {title}
  {\bibinfo {title} {High quality factor phase gradient metasurfaces},\
  }\href@noop {} {\bibfield  {journal} {\bibinfo  {journal} {Nature
  Nanotechnology}\ }\textbf {\bibinfo {volume} {15}},\ \bibinfo {pages} {956}
  (\bibinfo {year} {2020})}\BibitemShut {NoStop}%
\bibitem [{\citenamefont {Hu}\ \emph {et~al.}(2018)\citenamefont {Hu},
  \citenamefont {Li}, \citenamefont {Tong}, \citenamefont {Wu}, \citenamefont
  {Xia}, \citenamefont {Wang}, \citenamefont {Li}, \citenamefont {Huang},
  \citenamefont {Wang}, \citenamefont {Hou} \emph {et~al.}}]{hu2018type}%
  \BibitemOpen
  \bibfield  {author} {\bibinfo {author} {\bibfnamefont {C.}~\bibnamefont
  {Hu}}, \bibinfo {author} {\bibfnamefont {Z.}~\bibnamefont {Li}}, \bibinfo
  {author} {\bibfnamefont {R.}~\bibnamefont {Tong}}, \bibinfo {author}
  {\bibfnamefont {X.}~\bibnamefont {Wu}}, \bibinfo {author} {\bibfnamefont
  {Z.}~\bibnamefont {Xia}}, \bibinfo {author} {\bibfnamefont {L.}~\bibnamefont
  {Wang}}, \bibinfo {author} {\bibfnamefont {S.}~\bibnamefont {Li}}, \bibinfo
  {author} {\bibfnamefont {Y.}~\bibnamefont {Huang}}, \bibinfo {author}
  {\bibfnamefont {S.}~\bibnamefont {Wang}}, \bibinfo {author} {\bibfnamefont
  {B.}~\bibnamefont {Hou}}, \emph {et~al.},\ }\bibfield  {title} {\bibinfo
  {title} {Type-ii dirac photons at metasurfaces},\ }\href@noop {} {\bibfield
  {journal} {\bibinfo  {journal} {Physical Review Letters}\ }\textbf {\bibinfo
  {volume} {121}},\ \bibinfo {pages} {024301} (\bibinfo {year}
  {2018})}\BibitemShut {NoStop}%
\bibitem [{\citenamefont {Vieu}\ \emph {et~al.}(2000)\citenamefont {Vieu},
  \citenamefont {Carcenac}, \citenamefont {Pepin}, \citenamefont {Chen},
  \citenamefont {Mejias}, \citenamefont {Lebib}, \citenamefont
  {Manin-Ferlazzo}, \citenamefont {Couraud},\ and\ \citenamefont
  {Launois}}]{vieu2000electron}%
  \BibitemOpen
  \bibfield  {author} {\bibinfo {author} {\bibfnamefont {C.}~\bibnamefont
  {Vieu}}, \bibinfo {author} {\bibfnamefont {F.}~\bibnamefont {Carcenac}},
  \bibinfo {author} {\bibfnamefont {A.}~\bibnamefont {Pepin}}, \bibinfo
  {author} {\bibfnamefont {Y.}~\bibnamefont {Chen}}, \bibinfo {author}
  {\bibfnamefont {M.}~\bibnamefont {Mejias}}, \bibinfo {author} {\bibfnamefont
  {A.}~\bibnamefont {Lebib}}, \bibinfo {author} {\bibfnamefont
  {L.}~\bibnamefont {Manin-Ferlazzo}}, \bibinfo {author} {\bibfnamefont
  {L.}~\bibnamefont {Couraud}},\ and\ \bibinfo {author} {\bibfnamefont
  {H.}~\bibnamefont {Launois}},\ }\bibfield  {title} {\bibinfo {title}
  {Electron beam lithography: resolution limits and applications},\ }\href@noop
  {} {\bibfield  {journal} {\bibinfo  {journal} {Applied Surface Science}\
  }\textbf {\bibinfo {volume} {164}},\ \bibinfo {pages} {111} (\bibinfo {year}
  {2000})}\BibitemShut {NoStop}%
\end{thebibliography}

\begin{thebibliography}{1}%
\makeatletter
\providecommand \@ifxundefined [1]{%
 \@ifx{#1\undefined}
}%
\providecommand \@ifnum [1]{%
 \ifnum #1\expandafter \@firstoftwo
 \else \expandafter \@secondoftwo
 \fi
}%
\providecommand \@ifx [1]{%
 \ifx #1\expandafter \@firstoftwo
 \else \expandafter \@secondoftwo
 \fi
}%
\providecommand \natexlab [1]{#1}%
\providecommand \enquote  [1]{``#1''}%
\providecommand \bibnamefont  [1]{#1}%
\providecommand \bibfnamefont [1]{#1}%
\providecommand \citenamefont [1]{#1}%
\providecommand \href@noop [0]{\@secondoftwo}%
\providecommand \href [0]{\begingroup \@sanitize@url \@href}%
\providecommand \@href[1]{\@@startlink{#1}\@@href}%
\providecommand \@@href[1]{\endgroup#1\@@endlink}%
\providecommand \@sanitize@url [0]{\catcode `\\12\catcode `\$12\catcode
  `\&12\catcode `\#12\catcode `\^12\catcode `\_12\catcode `\%12\relax}%
\providecommand \@@startlink[1]{}%
\providecommand \@@endlink[0]{}%
\providecommand \url  [0]{\begingroup\@sanitize@url \@url }%
\providecommand \@url [1]{\endgroup\@href {#1}{\urlprefix }}%
\providecommand \urlprefix  [0]{URL }%
\providecommand \Eprint [0]{\href }%
\providecommand \doibase [0]{https://doi.org/}%
\providecommand \selectlanguage [0]{\@gobble}%
\providecommand \bibinfo  [0]{\@secondoftwo}%
\providecommand \bibfield  [0]{\@secondoftwo}%
\providecommand \translation [1]{[#1]}%
\providecommand \BibitemOpen [0]{}%
\providecommand \bibitemStop [0]{}%
\providecommand \bibitemNoStop [0]{.\EOS\space}%
\providecommand \EOS [0]{\spacefactor3000\relax}%
\providecommand \BibitemShut  [1]{\csname bibitem#1\endcsname}%
\let\auto@bib@innerbib\@empty
\bibitem [{\citenamefont {Gao}\ \emph {et~al.}(2022)\citenamefont {Gao},
  \citenamefont {Sain},\ and\ \citenamefont {Zentgraf}}]{gao2022spin}%
  \BibitemOpen
  \bibfield  {author} {\bibinfo {author} {\bibfnamefont {W.}~\bibnamefont
  {Gao}}, \bibinfo {author} {\bibfnamefont {B.}~\bibnamefont {Sain}},\ and\
  \bibinfo {author} {\bibfnamefont {T.}~\bibnamefont {Zentgraf}},\ }\bibfield
  {title} {\bibinfo {title} {Spin-orbit interaction of light enabled by
  negative coupling in high-quality-factor optical metasurfaces},\ }\href@noop
  {} {\bibfield  {journal} {\bibinfo  {journal} {Physical Review Applied}\
  }\textbf {\bibinfo {volume} {17}},\ \bibinfo {pages} {044022} (\bibinfo
  {year} {2022})}\BibitemShut {NoStop}%
\end{thebibliography}
%


\clearpage
\onecolumngrid
\begin{center}
  {\LARGE\bfseries Supporting Information\par}
  \vspace{1.5em}
\end{center}
\twocolumngrid

\renewcommand\thefigure{S\arabic{figure}} 
\setcounter{figure}{0}
\section{Sample fabrication}
\begin{figure}[h]
    \centering
    \includegraphics[width=1.0\columnwidth]{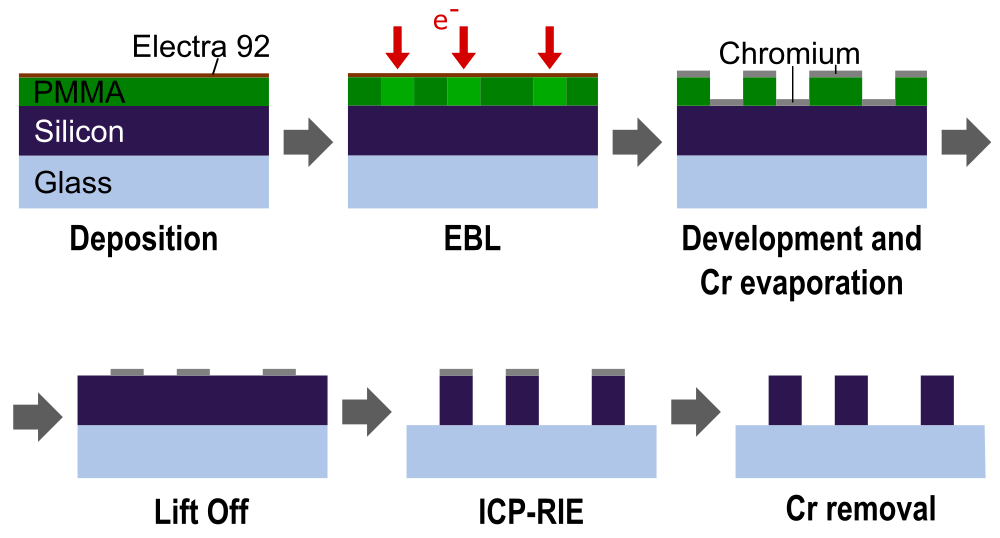}
\caption{Sketched sample fabrication steps starting with the plasma-enhanced chemical vapor deposition (PECVD) followed by the deposition and exposure of the electron beam lithography (EBL) positive resist PMMA. The developed PMMA acts as a mask for Chromium, which then, after the lift off,  acts as a mask for inductively coupled plasma reactive ion etching (ICP-RIE).}
    \label{fabrication}
\end{figure}

%
The metasurface is fabricated by following the steps shown in Figure \ref{fabrication}. First, the quartz-glass substrate is cleaned, and then amorphous silicon is deposited using plasma-enhanced chemical vapor deposition (PECVD). Then, the positive tone resist Polymethylmethacrylat (PMMA) is spin coated on top as well as a conductive layer of Electra 92.
The regions designated for the waveguides were exposed to an electron dose of \SI{450}{\micro\coulomb\per\square\centi\meter}, inducing scission of the PMMA molecules that increases solubility in the developer.
The electron beam lithography (EBL) was done using the \texttt{Raith Voyager EBL system}. After development with methyl isobutyl ketone (MIBK), \SI{15}{\nano\meter} of Chromium are evaporated on top of the PMMA Mask with a rate of \SI{1}{\angstrom\per\second}. 
A lift-off process removes the PMMA and reveals the chromium etching mask for the subsequent plasma reactive ion etching (ICP-RIE). This etching removes the silicon that is not covered by the chromium anisotropically. A chemical removal of the chromium mask finishes the process.  A scanning electron micrograph of an exemplary metasurface is shown in Figure 3 of the main text.
%
\section {Experimental system}
\begin{figure}[h!]
    \centering    
    \includegraphics[width=\columnwidth]{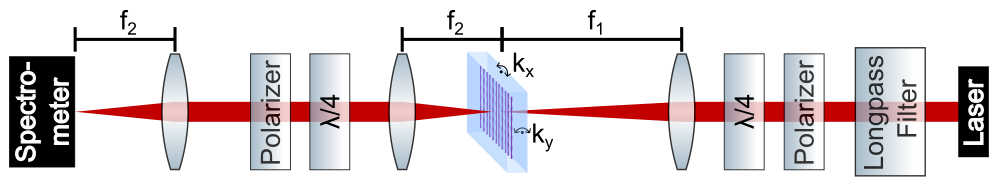}
\caption{Sketch of the optical setup for transmission measurements}
    \label{fig:setup}
\end{figure}

The experimental setup sketched in Figure \ref{fig:setup} is used to measure the transmission spectrum of the metasurface for multiple angle of incidence and polarizations. The light source is an unpolarized white-light \texttt{Fianium} laser with a pump-wavelength of \SI{1050}{\nano\meter} that is filtered out by longpass filter. The polarization state is then set by a broadband polarizer (and an additional quarter-waveplate for circular polarization). The beam is focused onto the sample with focus length of \SI{250}{\milli\meter}. This focal length is a good trade-off between the size of the metasurface (\SI{400}{}x\SI{400}{\micro\meter}) and the spread of the wavevector $k$. The transmitted light is collimated by a $f_2=\SI{100}{\milli\meter}$ lens, analyzed in polarization (for linear polarization measurements the quarter-waveplate is removed) and focused onto a \texttt{Shamrock} Spectrometer with grating period $300~\frac{l}{\mathrm{mm}}$. The sample is mounted onto a 3-axis linear stage combined with two rotational stages to cover multiple angles of incidence. 
\section {Resonance sharpness}
\begin{figure}[b]
\centering
\includegraphics[width=\columnwidth]{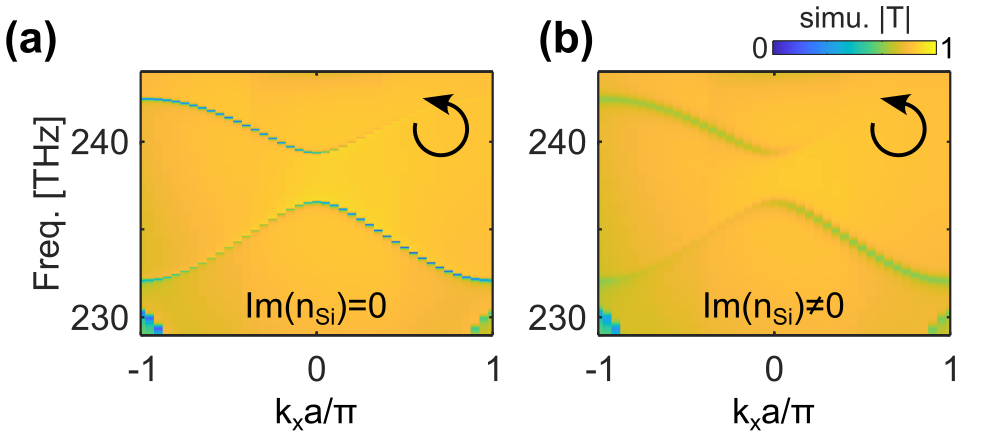}
\caption{Simulated transmission of a sweep in $k_x$ at $k_y=0.3\pi/b$ for LCP and $\Delta W=\SI{3.5}{\nano\meter}$ for (a)~ real refractive index of silicon (b)~complex refractive index of silicon measured by ellipsometry.}
    \label{Damping_by_losses}
\end{figure} 
The experiments discussed in the main text prove the existence of the simulated resonances. However, compared to simulated transmission, the observed resonances are much more diffuse and less sharp than the simulated ones. There are mainly two effects in the real experiment, that broaden the resonances and with that also decrease their depth. 

The major effect is due to losses in silicon. Although, these losses are quite low for the relevant wavelengths around \SI{1250}{\nano\meter} ($n_\mathrm{Si}=3.45+0.0034i$ with $n$ being the refractive index) and are therefore often neglected, 
they strongly impact the resonance in this metasurface. 
To show this damping effect, a simulation with real and complex refractive index are compared in Figure \ref{Damping_by_losses}. One can clearly see, that the resonances are less sharp and deep when losses in silicon are considered.

The second, less striking, damping effect results from the focusing of the beam, which leads to a wider range of $k$-vectors. Because the resonance frequency depends on the $k$-vector, multiple transmission dips of slightly different frequencies overlap leading to a damping of the observed resonance. This effect is  minimized by focusing onto the sample with a rather large focal length.
%
%
\section{Detailed data Analysis for $k_x=0$}
For $k_x=0$, i.e.~the analysis of the linear polarization eigenstates, the overall increase in energy with $k_y$ is present according to the Type-II Dirac Point. Therefore, the difference between the horizontally and vertically polarized eigenstates is difficult to visualize since the energy difference is way smaller than the overall energy sweep within the discussed range of $k_y$. Therefore, we chose a way to depict both modes in a single plot: namely imaging the total spectra for horizontal polarization and the resonance energy for vertical polarization by an additional pink line. 

\begin{figure}[]
    \centering
\includegraphics[width=1.0\columnwidth]{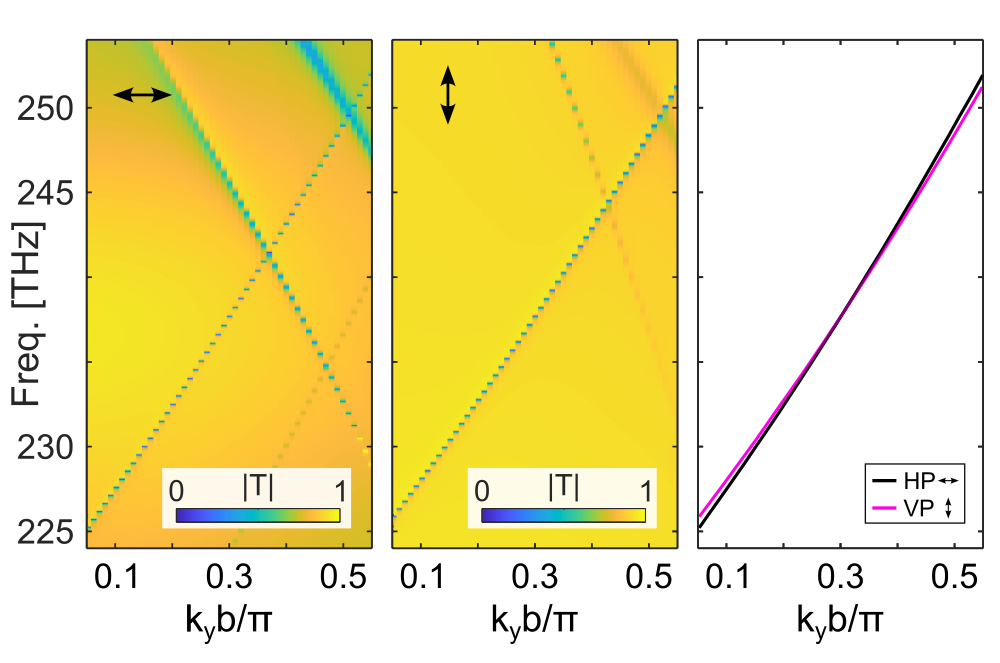}
\caption{Simulated transmission of the metasurface for a sweep in $k_y$ at $k_x=0$ for horizontally (left) and vertically (center) polarized incident light and $\Delta W=0$. The two lines on the right follow the energies of the simulated resonances.  }
    \label{kx0_fig2}
\end{figure}

Figure \ref{kx0_fig2} shows the extended 
data of Figure 2(d) in the main text, i.e.~the simulated transmission spectra for linear horizontal polarization (left) and vertical polarization (center) with $\Delta W=0$. The right diagram summarizes the resonance energies for both linear polarization, reveling the crossing of the modes. 

\begin{figure}[]
    \centering
    \includegraphics[width=1.0\columnwidth]{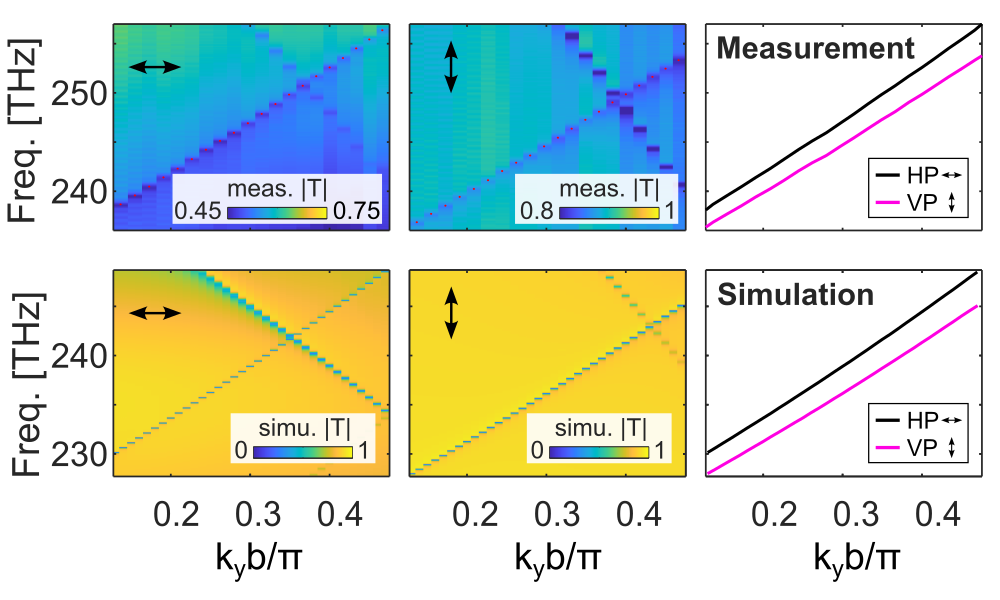}
\caption{Measured (top) and simulated (bottom) transmission of the metasurface for a sweep in $k_y$ at $k_x=0$ for horizontally (left) and vertically (center) polarized incident light. The two lines on the right follow the energies of resonances present in the transmission spectrum. For the simulations (bottom) $\Delta W=3.5$ was taken into account. }
    \label{kx0_fig4}
\end{figure}

The comprehensive data for Figure 4(e) and (f) are shown in Figure \ref{kx0_fig4} at the top and bottom, respectively, whereas the top shows measured data and the bottom simulated ones. Like in the previous Figure, the three columns correspond to horizontal polarization (left), vertical polarization (center) and summarized resonance energies (right). The similarity between measurement and simulation is striking, as well as the absence of mode crossing.
%
%
\section {Effect of width offset}

To cover the effect of a waveguide width offset in our tight-binding model, we extend the Hamiltonian derived by Gao et al.~\cite{gao2022spin} by an additional term to change the description from a ($\delta k_x$,$\delta k_y$) region around the Dirac point towards a point ($K_x$,$K_y$) in $k$-space to describe its absolute position. The extended tight-binding model reads:

\begin{equation}
    \mathcal{H}_{\text{eff}} = \mathcal{H}_{QQ}-\frac{i}{2}\Gamma\Gamma^\dagger+\mathcal{H}_{\text{extra}}-\mathcal{H}_{\text{shift}},
    \label{eq:tight}
\end{equation}
where
\begin{equation}
    \mathcal{H}_{QQ} = T\delta k_yI+v_y\delta k_y\sigma_z+v_x\delta k_x\sigma_y
\end{equation}
is the basic tight-binding Hamiltonian,
\begin{equation}
    \Gamma^\dagger = \begin{pmatrix}
        \kappa & \eta e^{i(\gamma+\Delta)}\delta k_x/k_d\\
        -\kappa e^{i\gamma}\delta k_x/k_d & \eta e^{i\Delta}
    \end{pmatrix}
\end{equation}
that dictates the coupling to the ambient environment.
\begin{equation}
    \mathcal{H}_{\text{extra}} = i(-v_t\delta k_yI+v_s\delta k_y\sigma_z)
\end{equation}
incorporates the imaginary dispersion in $\delta k_y$ and
\begin{equation}
    \mathcal{H}_{\text{shift}} = (T-iv_t)K_yI+(v_y+iv_s)K_y\sigma_z+v_xK_x\sigma_y
\end{equation}
shifts the Dirac point to $(K_x,K_y)$. We solve the respective eigenvalue problem and calculate the degree of circular polarization $\sigma_d = -2\frac{\text{Im}(v_1v_2^*)}{|v_1|^2+|v_2|^2}$ for the eigenvectors $v=(v_1, v_2)^\mathrm{T}$. To reproduce the experimental data the parameters $v_x,v_y,K_y$ were fitted to simulated data of eigenenergies in the form $\frac{1}{2}\text{Re}(E_2-E_1)$. The fit provides the parameter values $v_x = (5.25\pm0.09)10^{6}~\mathrm{Hzm}$, $v_y = (0.5\pm 0.2)10^{6}~\mathrm{Hzm}$, and $K_yb/\pi = 0.453\pm0.009$. $K_x$ was not fitted, since the DP is known to lie entirely at $k_x = 0$. The other parameters of the
\begin{figure}[h!]
    \centering
    \includegraphics[width=1.0\columnwidth]{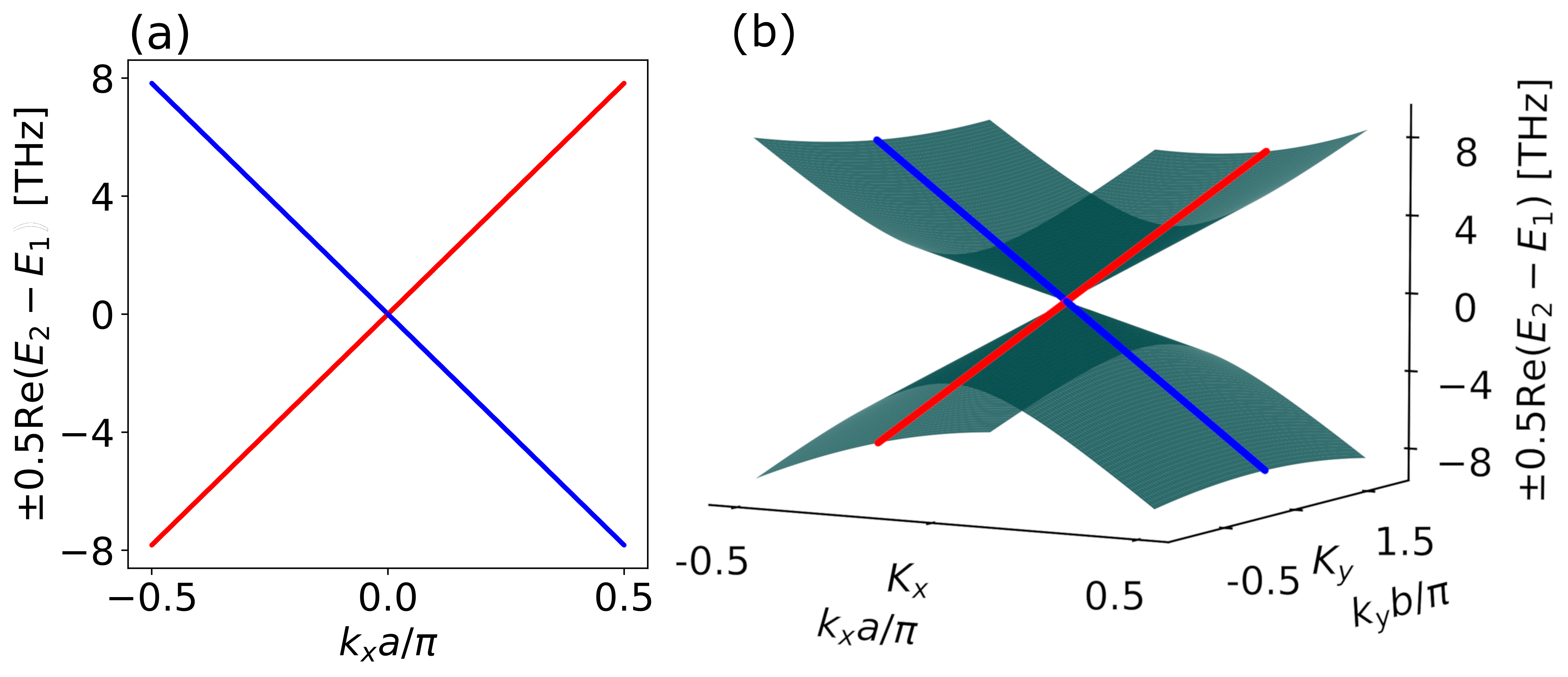}
\caption{Tight-binding model calculations demonstrating the shifted Dirac point at ($K_x=0, K_y=0.46$) in reciprocal space. (a) Eigenvalue crossing at the Dirac point with right- (blue) and left- (red) circularly polarized eigenmodes forming the singularity. (b) Eigenvalue surface of the shifted Dirac cone together with the lines shown in panel (a).}
    \label{fig:tbmodel}
\end{figure}
model not mentioned above are taken from the supplement of \cite{gao2022spin} where they were retrieved from numerical simulations of a similar metasurface.

In Figure~\ref{fig:tbmodel}, the resulting eigenvalues around the Dirac point are depicted. In Figure~\ref{fig:tbmodel}(a) the eigenvalues along $k_yb/\pi=0.46$ are shown. In agreement with the experiment we retrieve one left and one right circularly polarized eigenstate depicted in red and blue respectively, crossing at the Dirac point. In figure~\ref{fig:tbmodel}(b) the eigenvalue surface in 
\begin{figure}[h!]
    \centering
    \includegraphics[width=1.0\columnwidth]{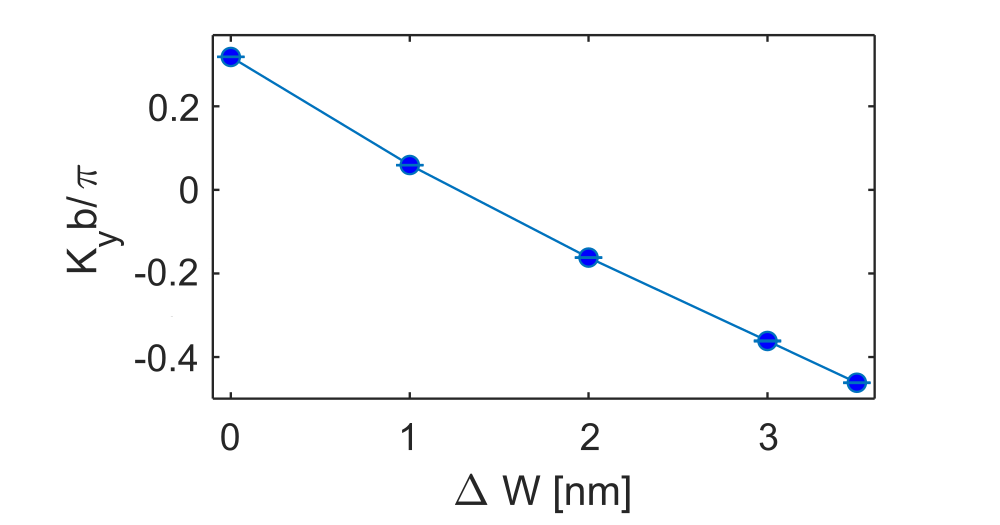}
\caption{$k_y$ Position of the Dirac point in $K_y$ shifting with the width offset $\Delta W$.}
   \label{suppl_fig:6}
\end{figure} 
the vicinity of the Dirac point is depicted together with the lines 
from panel (a) to illustrate the degeneracy shifting. Here, the eigenenergies $E_1,E_2$ of the effective Hamiltonian in eq.~(\ref{eq:tight}) were calculated and plotted as $\pm\frac{1}{2}\text{Re}(E_2-E_1)$, effectively modeling a Dirac cone with zero tilt. The results are in good agreement with the Maxwell simulations and up to a shift with the experimental data shown in Figure~5 in the main part of our manuscript.

Based on this, the Dirac point can be localized in parameter space in dependence of the width offset. In Figure~\ref{suppl_fig:6} the shift of the Dirac point in $k_y$-direction is shown for increasing waveguide width offset $\Delta W$.
%
%

%

%
%
\end{document}